\documentclass[aps,pre,twocolumn,superscriptaddress,showpacs,showkeys]{revtex4-2}
\usepackage{graphicx}                   
\usepackage{hyperref}                   
\usepackage{amsmath,amssymb}            
\usepackage{epstopdf}
\usepackage{subcaption}					
\usepackage{float}
\usepackage{color}
\definecolor{orange}{rgb}{1,0.5,0}
\definecolor{brown}{rgb}{0.65, 0.16, 0.16}
\definecolor{phlox}{rgb}{0.87, 0.0, 1.0}

\graphicspath{{figs/}}					

\begin{document}

	\title{Sandpiles Subjected to Sinusoidal Drive}
	
	\author{J. Cheraghalizadeh}
	\affiliation{Department of Physics, University of Mohaghegh Ardabili, P.O. Box 179, Ardabil, Iran}
	
	\author{M. A. Seifi MirJafarlou}
	\affiliation{Department of Physics, University of Mohaghegh Ardabili, P.O. Box 179, Ardabil, Iran}
	
	\author{M. N. Najafi}
	\affiliation{Department of Physics, University of Mohaghegh Ardabili, P.O. Box 179, Ardabil, Iran}
	\email{morteza.nattagh@gmail.com}

	\begin{abstract}
    This paper considers a sandpile model subjected to a sinusoidal external drive with the time period $T$. We develop a theoretical model for the Green function in a large $T$ limit, which predicts that the avalanches are anisotropic and elongated in the oscillation direction. We track the problem numerically and show that the system shows additionally a regime where the avalanches are elongated in the perpendicular direction with respect to the oscillations. We find a transition point between these two regimes. The power spectrum of avalanche size and the grains wasted from the parallel and perpendicular directions are studied. These functions show power-law behaviour in terms of the frequency with exponents, which run with $T$. 
	\end{abstract}

	\pacs{05., 05.20.-y, 05.10.Ln, 05.45.Df}
	\keywords{ Oscillating Sandpile, Self-organized criticality, Critical exponent, Scaling relation }
	\maketitle

	\section{Introduction}
	Sandpile models are prototypical examples of the systems which show self-organized criticality (SOC) that is self-sustaining in a critical state without tuning any external parameter. To cover various SOC systems found in nature, plenty of versions of sandpiles have been introduced so far, each of which manipulates an essential aspect of the original sandpile model introduced by Bak, Tang and Weisenfeld (BTW). The question of how the universality class of sandpile is changed fueled a lot of studies in the field. The Manna version of sandpiles (Manna model, sometimes called two-state sandpile model) introduces \textit{stochasticity} to the BTW model~\cite{manna1991two,dhar1999some}, while the Zhang model~\cite{lubeck1997large,chessa1999universality} for sandpiles makes it continuum. There are also directed versions of sandpiles that have been enormously investigated over time~\cite{pastor2000universality,dhar1999abelian,hasty1998renormalization}. The sandpile models has also been implemented on top of many host systems like the regular $d$-dimensional hypercubic lattice~\cite{dhar1999abelian,ben1996universality,najafi2016bak}, honeycomb lattice~\cite{najafi2012avalanche}, Bethe lattice~\cite{dhar1990abelian}, random network~\cite{najafi2014bak,fronczak2006self,bonabeau1995sandpile, redig2012abelian}, scale-free network~\cite{karmakar2005sandpile,lee2004sandpile}, small-world network~\cite{bhaumik2013critical,pan2007sandpile,bhaumik2018stochastic,hoore2013critical,najafi2018statistical,lahtinen2005sandpiles}, etc. Many aspects of this model is known, like the height correlations~\cite{majumdar1991height}, and its relation to the other statistical models, like the $q$-state Potts model and Spanning trees~\cite{majumdar1992equivalence,manna1992spanning}, loop-erased random walks~\cite{majumdar1992exact}, logarithmic conformal field theories~\cite{piroux2004pre,moghimi2005abelian,ruelle2013logarithmic}, and Schramm-Loewner evolution~\cite{najafi2012avalanche,saberi2009direct}. For review see~\cite{dhar1999some,najafi2021some,markovic2014power}.\\
	
	The real sandpiles were considered in a few studies. In~\cite{held1990experimental} the properties of real sandpiles on a circular disk were analyzed, where the sand grains were added to the pile after the avalanches subside. In this study, self-similar avalanches with power-law mass distribution function were observed, and a power spectrum of mass was found to behave like $f^{-2}$ ($f$ being the frequency). Jaeger \textit{et. al.}~\cite{jaeger1989relaxation} arranged an experimental setup for sandpiles consisting of spherical glass beads or rough aluminium-oxide particles, where the effect of vibration was also tested. For the case without vibrations, serious deviations from the theoretical predictions were found for the avalanche duration and the power spectrum of the system. It is believed that the deviations from the power-law behaviour for the power spectrum predicted by BTW are due to the hysteresis angle (the difference between the threshold angle above which avalanches form and the metastable angle). In fact, for each toppling, the metastable angle should be exceeded by this hysterics angle in order to allow sands to slide downhill, the fact that causes the periodic occurrence of avalanches in the steady state~\cite{bagnold2012physics}. The generation of vibrations (which were generated by connecting the system to speakers) as an external drive of the sandpile ~\cite{jaeger1989relaxation} has shown to have a drastic effect on the spectral properties of the system. Importantly it causes the system to show a self-similar power spectrum, which for the case of study in~\cite{jaeger1989relaxation} it is like $f^{-0.8}$, valid for high enough frequencies and independent of the type of particles. This problem was analytically followed by considering a periodic external drive. These observations raise an important question concerning the effect of vibrations on the universal behaviours of sandpiles: whether and how the vibrations modify the geometrical and local properties of sandpiles. The effect of vibrations and their properties have poorly been investigated in the literature, and some limited simulations have been done~\cite{mehta1991vibrated,barker1992vibrated,barker1993transient}. Despite the limitations of the theoretical modelling of sandpiles (the discrepancy between the exponents predicted by theoretical models and experiments), the theoretical modelling of the oscillating external drive on sandpile helps much to shed light on various aspects and expectations.\\
	
	In this study, we model the vibrating (periodic external drive) sandpiles with a constant rate of grain addition. To this end, we consider a pile subjected to periodic forcing. In the local toppling rule, the sand spread in the direction parallel to the vibrations and the perpendicular direction is different. We show that the power spectrum behaves in a power-law form, with the exponent which runs with the $T$ the period of the oscillations. We demonstrate that the system undergoes a crossover between two states: for small $T$ values, the avalanches are extended along the direction perpendicular to the vibrations. For large $T$ values, the avalanches are extended along the direction of the vibrations. To our best knowledge, this behaviour has not been noted before.\\
	
	The paper is organized as follows: We introduce the model in the next section. Section~\ref{SEC:results} is devoted to the results of the simulations. We close the paper with a conclusion.
	
	\section{The model}~\label{SEC:Model}
	
We define a sandpile model under oscillations (oscillatory sandpile model, OSM) on a $L\times L$ lattice and attribute the height variable $\left\lbrace h_{i,j}\right\rbrace_{i,j=1}^{L} $ to each site, which takes the values from the range $h_{i,j}\in \left[1,h_{\text{th}}\equiv4n \right] $ ($n$ being an integer number, which is 10 in this paper). When the height of any site $(i,j)$ exceeds the threshold $h_{\text{th}}$, it \textit{topples}, meaning that the site looses $h_{\text{th}}$ sand grains, which are distributed over its neighbors. More precisely, when the site $(i,j)$ topples, then $h_{i',j'}\rightarrow h_{i',j'}+\Delta_{i,j;i',j'}^t$, where $\Delta_{i,j;i',j'}^t$ is a time-dependent ($t$) matrix as follows: 
	\begin{equation}
		\Delta_{i,j;i',j'}^t=
		\begin{cases}
			\ \ \ \ \ \ -4n \ \ \ \ \ \ \ \ \ \  \  \ \text{if} \ i=i'\ \text{and} \  j=j'  \\
			\ \ \ \ \ \ \ \ \ n \ \ \ \  \ \ \ \ \ \ \ \  \ \text{if} \ i=i'\pm 1\ \text{and} \ j=j' \\
			[n(1\pm \epsilon_0 \sin(\omega t))] \  \text{if} \ i=i'\ \text{and} \ j=j'\pm 1 \\
			\ \ \ \ \ \ \ \ 0 \ \ \ \ \ \ \  \ \ \ \ \ \ \ \ \text{other }
		\end{cases}
		\label{Eq:toppling}
	\end{equation}
	\\
	where $\epsilon_0$ is the vibration strength parameter, $\omega = \frac{2\pi}{T}$ is the angular frequency, $[x]$ shows the integer value of $x$, and $T$ is the time period for the oscillations. Note that the oscillations are imposed only in the $j$ direction. Throughout this paper, we call the direction parallel to the direction of oscillations ($j$) as the \textit{parallel direction}, while the other direction (perpendicular to the oscillations, $i$) as the \textit{perpendicular} direction. Obviously, the OSM is anisotropic, i.e. the dynamics in the \textit{parallel} direction is different from the one in the \textit{perpendicular} ($i$) direction.
		\begin{figure}
		\centering
		\includegraphics[width=0.7\linewidth]{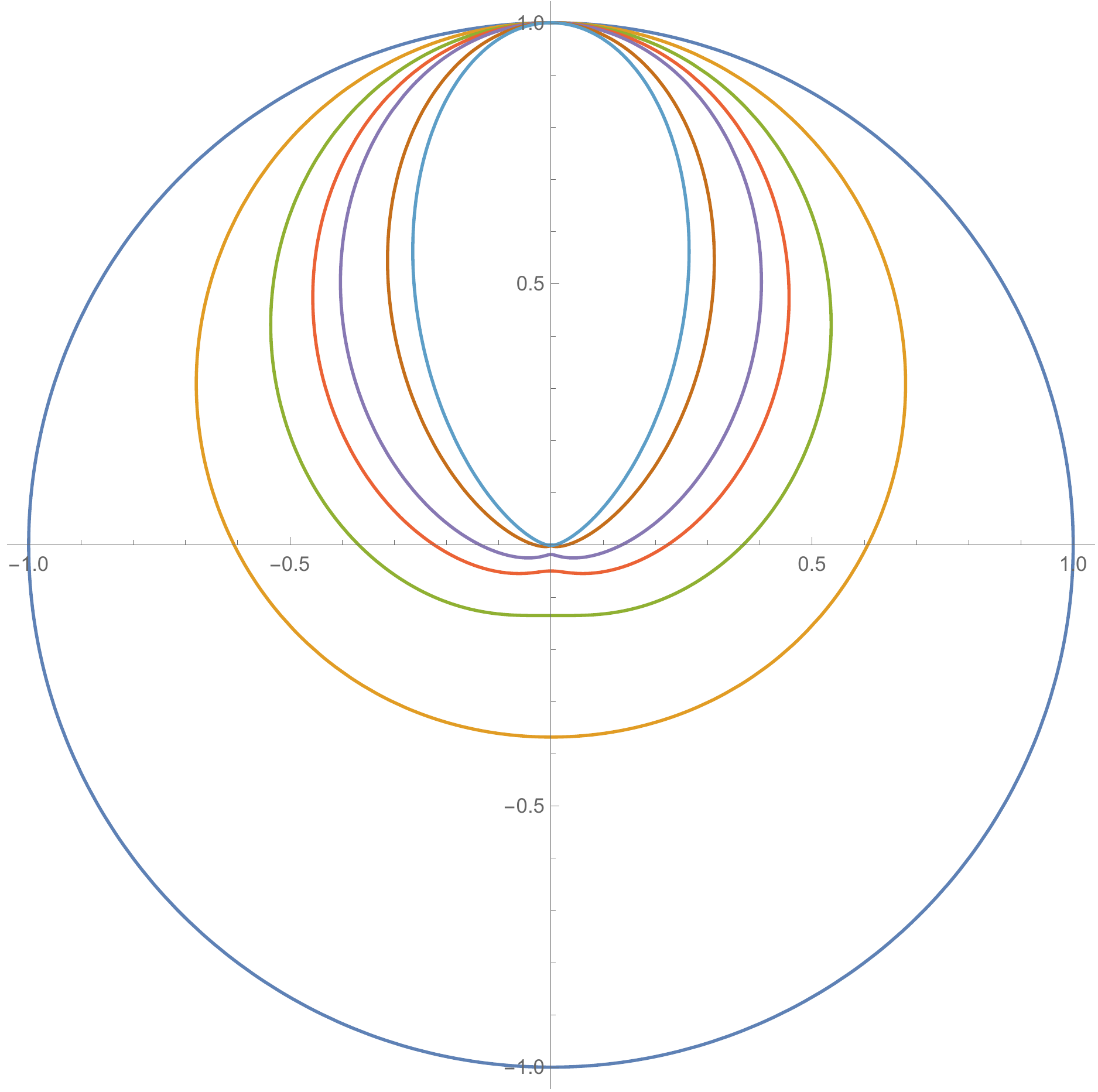}
		\caption{Polar plot for the asymptotic behavior $r\rightarrow\infty$ according to Eq.~\ref{Eq:asymptotic}. The symmetric circle path is for the isotropic case $\epsilon_0=0$, and the most anisotropic one (largest aspect ratio) is for $\epsilon_0=1$.}
		\label{Fig:polar}	
	\end{figure}
	\begin{figure*}
		\centering
		\includegraphics[width=1.0\linewidth]{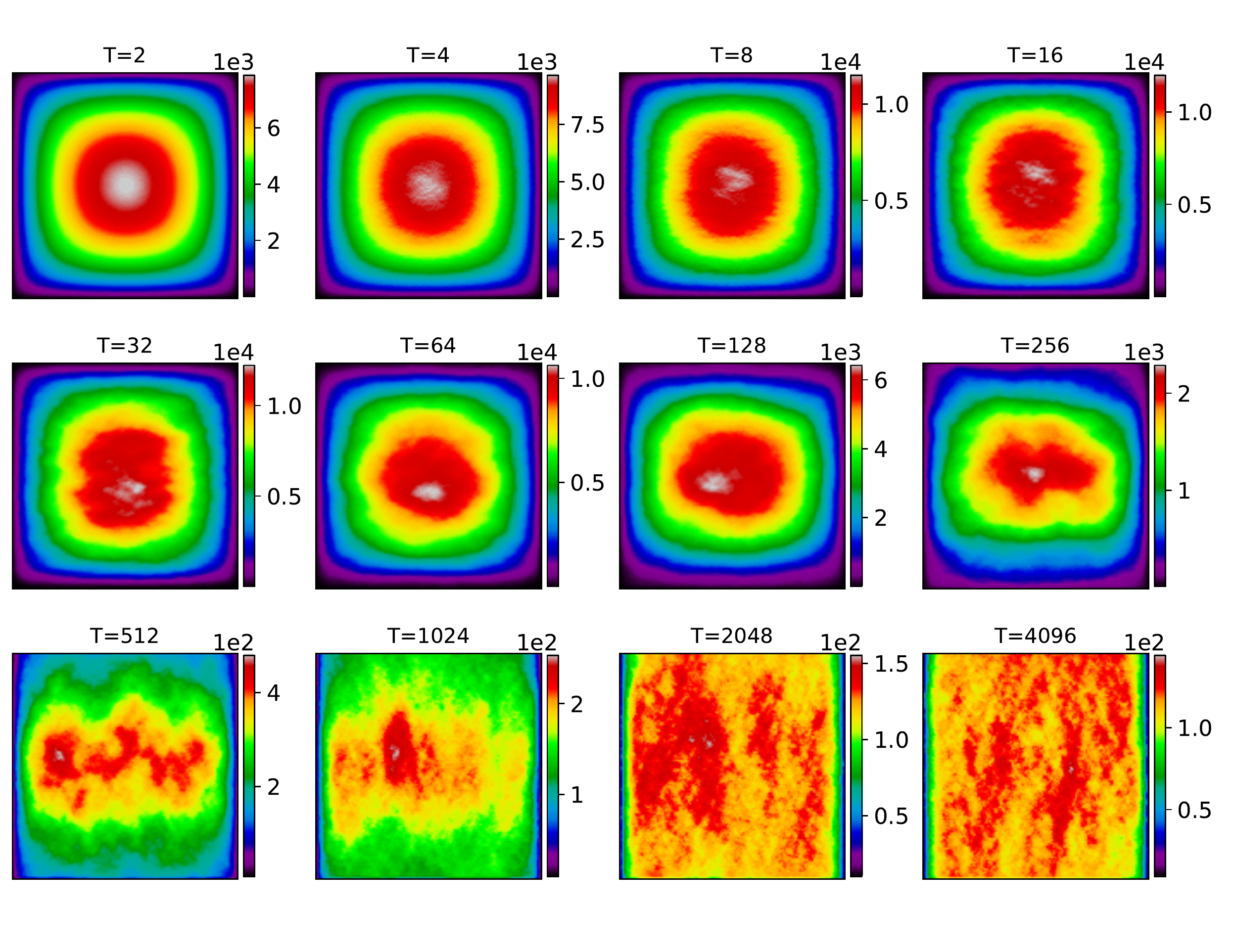}
		\caption{ Total system activity (sum of activity from $t=t_{\text{reached recurrent}}$ till $t=t_{\text{finished}}$) for various $T$ on $L=256$ lattice. The direction of oscillations (parallel direction) is up-down, and the perpendicular direction is left-right.}
		\label{Fig:total_act}	
	\end{figure*}
	
	Now we define the sandpile dynamic. We start with a uniform random configuration in which all the sites are stable. Then we add $n$ sand grains to a randomly chosen site. This operation, called \textit{grain injection} takes place every $\Delta t$ time step that is defined in the following. If it is stable again, then we wait for $\Delta t$ time steps to choose another random site for adding grains. If it becomes unstable, then it topples according to Eq.~\ref{Eq:toppling}. For the boundary sites, $n$ or $2n$ sand grains are dissipated depending on the position of the boundary site. As a result, the neighbouring sites may become unstable and topple in their turn. Therefore, a chain of topplings is triggered. We define \textit{one time step} $\delta t=1$ as the time during of which \textit{all} the sites of the lattice are tested once for local toppling, after which $t\rightarrow t+\delta t$. After $\Delta t$ time steps, another random site is chosen to add an external grain, whether or not the previous chain of activities is over: If the chain is over (there is no further unstable sites at some time), then the time is added one by one during which no activity is done until reaching the next time for injection (reaching $\Delta t$ steps). Therefore, the \textit{avalanches} are not well-defined as in the regular sandpiles (chain of topplings between two stable configurations). However, we have some \textit{toppled} sites in any arbitrary time interval $[t_1,t_2]$ which allows us to define avalanches: it is the set of sites that have toppled at least once in this interval where are \textit{unconnected avalanches}. All the geometrical statistical analyses are upon such avalanches in this paper.\\
	
	In the two limits $T=2$ and $T\rightarrow\infty$ the effect of oscillations is zero. When $T$ is high, in the short time scales, the sand grains tend to diffuse more in the parallel ($j$) direction, and the \textit{avalanches} are expected to extend in this direction. In the opposite limit, i.e. high $\omega$ values, one expects that the avalanche is squeezed in the parallel direction and extends more in the perpendicular direction. This is because the time scale for changing the direction of the preferred parallel direction (i.e. the direction to which the sand grains are more likely to topple according to Eq.~\ref{Eq:toppling}) due to the oscillations is much smaller than the time scale of avalanche to diffuse in this direction, and the avalanches do not have enough time to extend in the parallel direction. Therefore, we expect the avalanches to be anisotropic, and there is a transition point where the preferred alignment changes from the parallel to the perpendicular direction.\\
	
    We consider the following time series: The number of topplings at time $t$,  $a(t)$ which is called activity. The number of grains in the pile at time $t$, i.e. $m(t)$. These variables do not depend on the sign of bias, i.e. in terms of the Eq.~\ref{Eq:toppling} they would be a function of $\sin^2(wt)$ and have the periodicity $\frac{T}{2}$. The number of dissipated grains in the parallel (perpendicular) direction $dn_{\parallel}(t)$ ($dn_{\perp}(t)$) are the other quantities that we investigate here. These variables depend on the sign of the bias field, and they have the periodicity $T$. Note also that $m(t)=m(0)+\frac{t}{\Delta t}-\sum_{t'=1}^t \left[dn_{\parallel}(t')+ dn_{\perp}(t')\right] $, i.e. these quantities are not independent. The system's dynamics are divided into two parts: The first stages are called the transient configurations identified by the height increasing with time and configurations with nonzero probability in the periodic steady state. All of our analyses are done in the latter configurations. \\
		
	We start our arguments by defining the Green function $G(t,\textbf{X},\textbf{Y})$ as the probability that the site $\textbf{X}$ topples in an avalanche at time $t$ created by addition sand grain to the site $\textbf{Y}$ at $t=0$. The Green function satisfies the master equation ($\textbf{X}\ne \textbf{Y}$)
	\begin{widetext}
	\begin{equation}
		\begin{split}
			G(t+1,\textbf{X},\textbf{Y}) = \frac{1}{4} & \left[G(t,\textbf{X}+\textbf{e}_x,\textbf{Y})+G(t,\textbf{X}-\textbf{e}_x,\textbf{Y})\right.  \\
			& \left. +(1-\epsilon_0\sin(\omega t))G(t,\textbf{X}+\textbf{e}_y,\textbf{Y})+(1+\epsilon_0\sin(\omega t))G(t,\textbf{X}-\textbf{e}_y,\textbf{Y})\right].
		\end{split}
	\end{equation}
    \end{widetext}
	
	In the continuum limit one obtains ($\textbf{X}\ne\textbf{Y}$)
	\begin{equation}
	\begin{split}
	\partial_t G(t,\textbf{X},\textbf{Y}) & =D_2 \nabla^2G(t,\textbf{X},\textbf{Y})\\
	& -D_1\epsilon_0\sin \omega t\partial_y G(t,\textbf{X},\textbf{Y}),
	\end{split}
    \label{Eq:ScalingLimit}
	\end{equation}
	where $D_1=\frac{a}{2\delta t}$ and $D_2=\frac{a^2}{4\delta t}$ are the diffusion coefficients in the scaling limit, and $a$ is the lattice constant, and $\delta t$ is the time difference.
	Note that for the case $\epsilon_0=0$, it reduces to a simple diffusion equation. The solution of the above equation is not simple to find. Instead, let us consider simpler cases. The OSM with $\epsilon_0=1$ is mapped to a superposition of two \textit{deepest descent sandpile model (DDSM)} with preferred directions $\left\lbrace \text{nn}\right\rbrace_1 = \left\lbrace \textbf{e}_x,\textbf{e}_y\right\rbrace $ and $\left\lbrace \text{nn}\right\rbrace_2 =\left\lbrace \textbf{e}'_x,\textbf{e}_y\right\rbrace $ (where $\left\lbrace \text{nn}\right\rbrace$ shows the list of nearest neighbors, $\textbf{e}_x\equiv (1, 0)$, $\textbf{e}_y\equiv (0, -1)$, and $\textbf{e}'_x\equiv (-1, 0)$) just at the moments $\omega t_n=(2n+1)\frac{\pi}{2}$ ($n$ being an integer number). The DDSM model was investigated analytically in~\cite{dhar1989exactly}. For the other times, OSM is equivalent to a variant of this model where the diffusion in one preferred direction is stronger than the other direction. Although a complete mapping is restricted to the times $t_n$, some general features of OSM can be potentially understood in terms of DDSM (is easily mapped to a voter model), at least intuitively. Evidently, our model is more complicated since it is directed just in one direction, in contrast to the voter model for which there are two preferred directions. In the large $T$ limit, during a long time of dynamics the sand propagation in one direction (up or down) in much larger than the opposite direction. In this case $\sin \omega t$ is close to (positive or negative) unity for a long time, allowing us to talk about the steady state Green function which is independent of time, i.e. $g(\textbf{X})\equiv G(t,\textbf{X},\textbf{0})$. In this time interval the system is mapped effectively a DDSM, with directed diffusion in the $\textbf{e}_y$ direction, and the Eq.~\ref{Eq:ScalingLimit} becomes ($\textbf{X}\equiv (x,y)$, and $\sin \omega t\approx 1$)
	\begin{equation}
		D_2\left[\partial_x^2+\left(\partial_y-\kappa \right)^2- \kappa^2 \right] g(\textbf{X})=\delta^2(\textbf{X})
	\end{equation}
	where $\kappa^{-1}\equiv 2\frac{D_2}{D_1\epsilon_0}=a/\epsilon_0$ is proportional to the lattice constant. For the case $\textbf{X}\ne \textbf{0}$, the solution is 
	\begin{equation}
		g(\textbf{X})=e^{\kappa y}\left[ c_1 I_0\left( \kappa r\right) +c_2K_0\left(\kappa r\right)\right]
		\label{Eq:solution} 
	\end{equation} 
	where $r\equiv \sqrt{x^2+y^2}$, $I_n(z)$ and $K_n(z)$ are the modified Bessel function of the first and second kind, and $c_1$ and $c_2$ are constants to be determined. Since this solution is valid during the times smaller or in order to $T/2$, one should not be concerned about exponential growth in the positive $y$ direction. Noting that in the large $r$ limit, $I_0(\kappa r)\rightarrow (1/\sqrt{2\pi \kappa r}\exp\left[ \kappa r\right] )$ and $K_0(\kappa r)\rightarrow (\sqrt{\pi/2 \kappa r}\exp\left[ -\kappa r\right] )$, one concludes that $c_1=0$ and $c_2$ is non-zero. 
	
	Note that in the absence of oscillations $\epsilon_0=0$ or $\kappa=0$, the system is isotropic and $g$ depends only on $r$, so that $\frac{1}{r}\frac{\text{d}}{\text{d}r}\left[r \frac{\text{d}}{\text{d}r}g_{\epsilon_0=0}(r)\right]=\delta(r) $, the solution of which is $\ln r$ as a well-known fact in 2D ASM. 	Asymptotically, the Green function behaves like the following ($\textbf{X}\equiv (r,\theta)$, where $\theta =\tan^{-1}\left(\frac{y}{x}\right)$)
	\begin{equation}
		\begin{split}
			g(r,\theta)\rightarrow \left\lbrace \begin{matrix}
				e^{\kappa r\cos\theta}\left[ \text{const.}-\ln \frac{\kappa r}{2}\right]  & \kappa r\ll 1 \\
				\frac{1}{\sqrt{\kappa r}} \exp\left[ -\kappa r\left(1-\cos\theta \right)\right] & \kappa r\gg 1 		
			\end{matrix}\right.
		\end{split}
	\label{Eq:asymptotic}
	\end{equation}
The plots for the lower branch have been shown in Fig.~\ref{Fig:polar} for a constant $r$. Increasing $\epsilon_0$ from zero (isotropic case) makes the green function more oriented along the $y$ axis as expected. We see that the upper branch in Eq.~\ref{Eq:asymptotic} gives the known logarithmic result for $\epsilon_0=0$ ($\kappa=0$), i.e. the isotropic case. 
	A useful representation of Eq.~\ref{Eq:solution} is as follows ($c_1=0$)
	\begin{equation}
	g(r,\theta)\propto \int_r^{\infty}\frac{e^{-\kappa (R-r\cos \theta)}}{\sqrt{R^2-r^2}}dR
	\label{Eq:greenfunction}
	\end{equation}
	To interpret this equation, we take the strategy of Ref.\cite{ktitarev2000scaling}. Let us define $p(R,\theta)$ as the probability that the linear extent of an avalanche is $R$ in the $\theta$ direction and $\rho_R(r,\theta)$ as the density of sites (at $(r,\theta)$), covered by the avalanche with the extension $R$. Then it is easy to see that
	\begin{equation}
	g(r,\theta)\propto\int_r^{\infty} p(R,\theta)\rho_R(r,\theta)dR
	\end{equation}
	It was previously shown that for the isotropic case $p^{\epsilon_0=0}(R)\propto R^{-\tau_R}$ ($\tau_R=\frac{1}{4}$) and $\rho_R^{\epsilon_0=0}(r)\propto r^{d_f-2}=r^{-\frac{3}{4}}$, leading to $g_{\epsilon_0=0}(r)\propto \ln r$, see~\cite{ktitarev2000scaling,dashti2015statistical} for the details. In our case, to be consistent with Eq.~\ref{Eq:asymptotic}, and also with the isotropic $\epsilon_0=0$ case, we propose that ($x\equiv \frac{R}{r}$)
	\begin{equation}
	\begin{split}
	p(R,\theta) & \propto \frac{e^{-\kappa R(1-\cos\theta)}}{R^{\tau_R}} \\
	\rho_R(r,\theta) & \propto r^{d_f-2}\frac{e^{-\kappa r(x-1)\cos\theta}}{\sqrt{x^{\frac{3}{2}}-x^{-\frac{1}{2}}}}
	\end{split}
    \label{Eq:analyticalResults}
	\end{equation} 
	which results to Eq.~\ref{Eq:greenfunction}. Note that $P(R)$ has the same angular structure as the lower branch of Eq.~\ref{Eq:asymptotic}, see Fig.~\ref{Fig:polar}. These analytical arguments predict the directional deformation of the avalanches in the same form as the green function.\\
	
	Now we consider the system with a periodic external drive. For this case, the above equation is expected for long $T$ values and also times that the external drive is only in one direction, while for generic $t$ and $T$ the behaviour is much different. In fact, the system's response to the external drive is determined by comparing $t$ and $T$, and therefore, we need dynamical scaling arguments. For the DDSM, the probability that in the SOC state the avalanche has a duration greater than $t$ varies as $t^{-\frac{1}{2}}$ for large enough $t$ values~\cite{dhar1989exactly}. The same scaling relation is expected for our directed sandpile model, although the change of this scaling relation does not hurt our following arguments. This power-law behaviour should be modulated for a finite system with a finite-size effect function, leading to $\bar{D}\equiv \left\langle D \right\rangle\propto L^{\alpha_D}$ for the average duration of avalanches, where $\alpha_D$ is the corresponding exponent. Our arguments in the followings are based on the existence of such a finite time scale, not its exact form. To add the effect of changing the direction of the preferred orientation due to the oscillations, we approximate the sinusoidal function by a periodic step function $f(t)=+\epsilon_0$ ($=-\epsilon_0$) for $0\le t<T/2$ ($-T/2\le t<0$) with a period $T$. As a consequence, the probability that an avalanche experiences a sign change of the preferred direction $P(D>T)$ is proportional to $T^{-\frac{1}{2}}$. One then expects two distinct regimes: $T\ll \bar{D}$ and $T\gg \bar{D}$. For the former case, it is easy to show that $P(D>T)$ is considerably large in the thermodynamic limit, while for the latter case, it is negligibly small. Therefore, for the former case, the avalanches are squeezed in the parallel direction and are more extended in the perpendicular direction since they do not have enough time to completely settle/be established in the parallel direction. For the latter case, the avalanches would not experience this sign change and are more extended in the parallel direction.  \\

We consider the auto-correlation function (ACF) and the power spectral density (PSD) for these time series associated with the quantities introduced above. Given the measurements $\left\lbrace t_i,x_i\right\rbrace_{i=1}^N$ where  $t_i$ are times, the ACF is defined as
	\begin{equation}
		C_x(\tau)= \frac{\sum_{i=1}^{N}[(x(t_i)-\bar{x})(x(t_i+\tau)-\bar{x})]}{\sum_{i=1}^{N}(x(t_i)-\bar{x})^2},
	\end{equation}
	where  $x\in \{a, m, dv, dp \}$, and $\bar{x}$ shows average of time series, and $\tau$ is some time lag. The PSD is the Fourier transform of ACF, i.e. (the increment of $\tau$ is one, i.e. $\Delta \tau=1$)
	\begin{equation}
		\begin{split}
			A_x(f) \equiv \sum_{i=1}^{N} C_x(\tau_i) e^{-i2\pi f \tau_i} ,
		\end{split}
	\end{equation}
	where $f$ is frequency.

\section {Results and Analysis}\label{SEC:results}
In this section, we present the results of simulating OSM. Throughout this section we set $\epsilon_0 =1.0$, $n=10$, and $\Delta t=100$. Figure~\ref{Fig:total_act} shows the total (accumulated) activity of the system is shown in terms of $T$ (note that the case $T=2$ is isotropic as expected from Eq.~\ref{Eq:toppling}). This helps to figure out how the arguments presented in the previous section work, i.e. for small $T$ values, the avalanches are spatially extended in the perpendicular direction. In contrast, for large $T$ values, their extension in the parallel direction is more than the other direction. Figure~\ref{Fig:total_act_contour_1000} in Appendix \ref{SEC:AppendixA} shows a same effect for the avalanches.

\begin{figure}
 	\begin{subfigure}[b]{0.32\linewidth}
	 	\includegraphics[width=\linewidth]{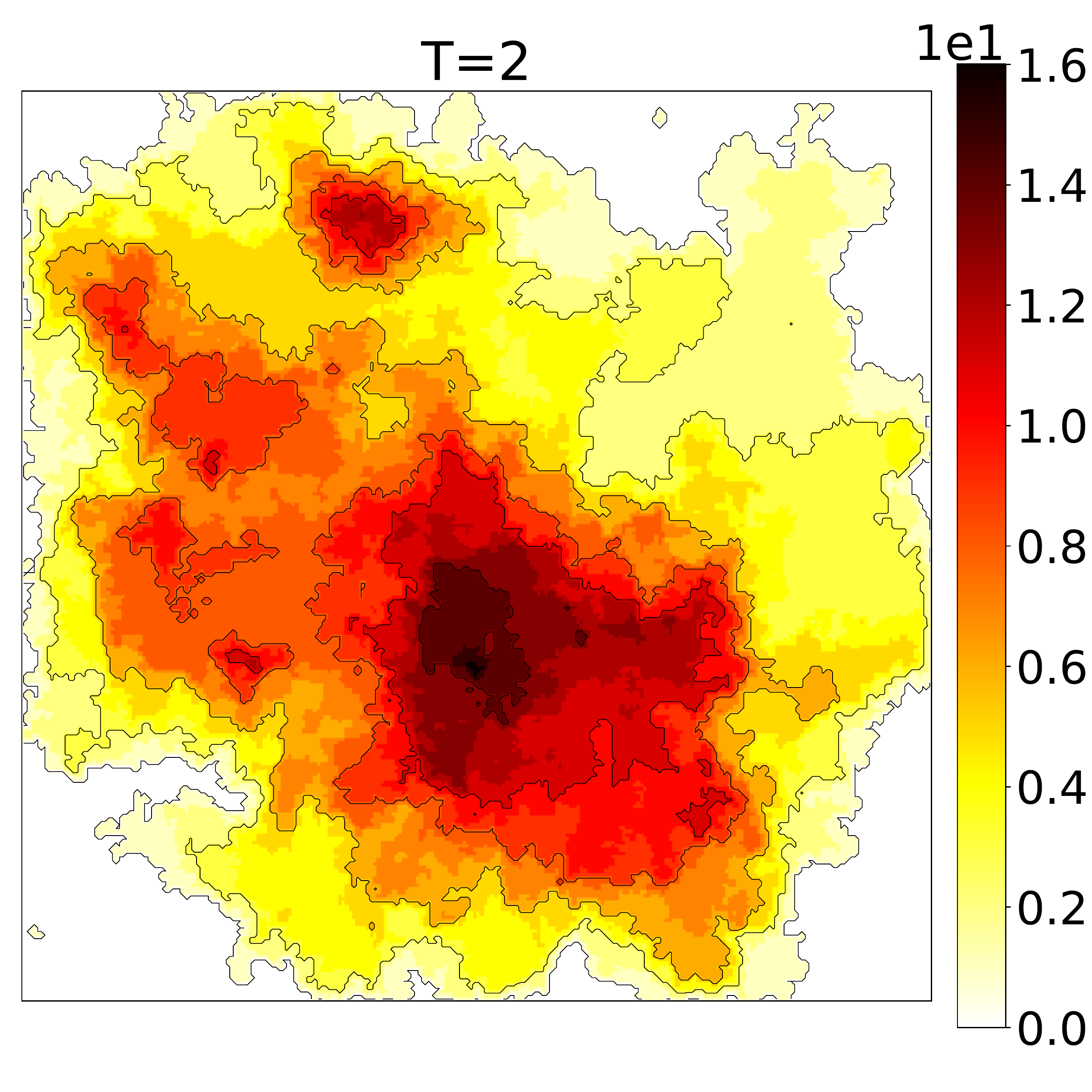}
	 	\caption{}
	 	\label{Fig:total_act_contour1}
	 \end{subfigure}
	 \begin{subfigure}[b]{0.32\linewidth}
	 	\includegraphics[width=\linewidth]{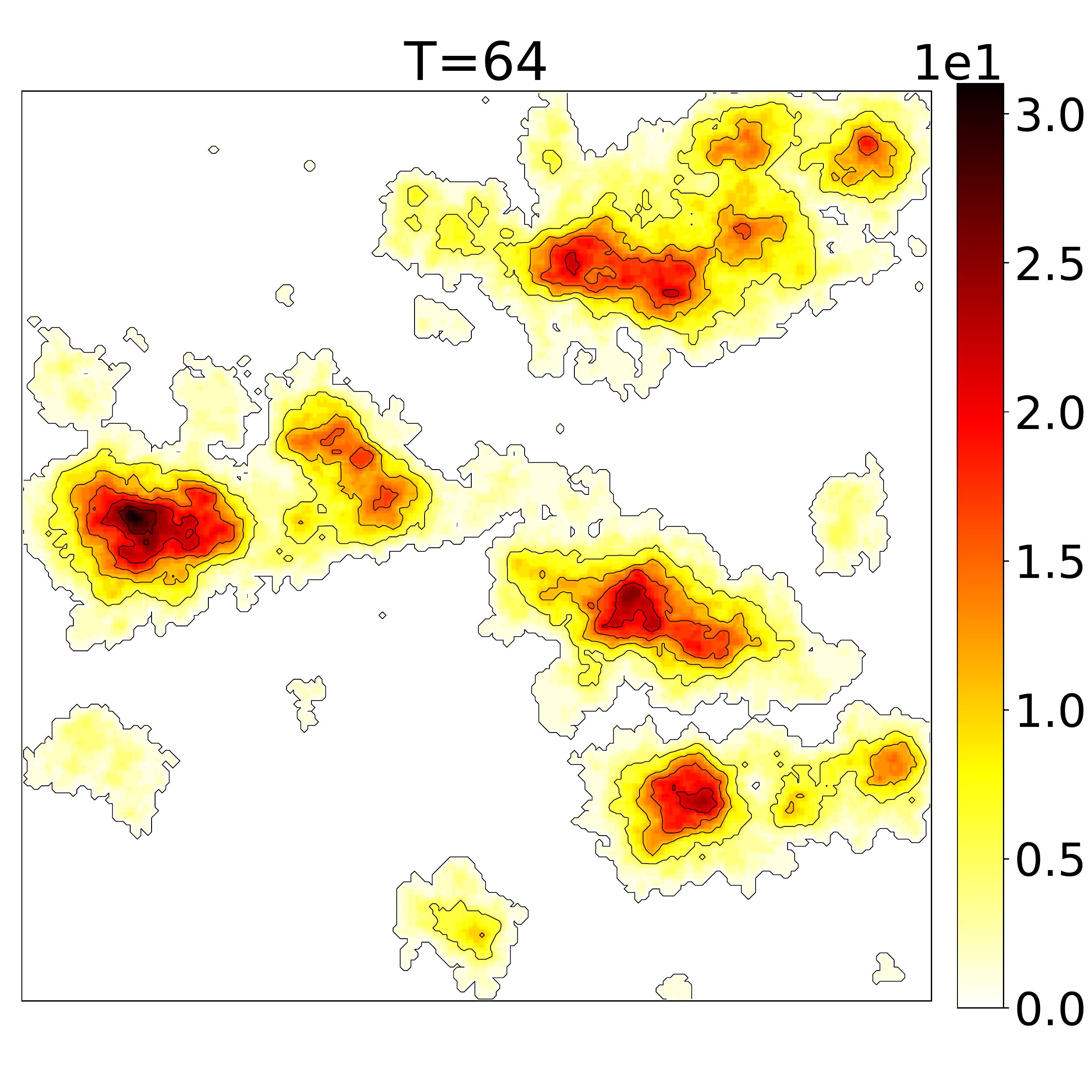}
	 	\caption{}
	 	\label{Fig:total_act_contour2}
	 \end{subfigure}
	 \begin{subfigure}[b]{0.32\linewidth}
	 	\includegraphics[width=\linewidth]{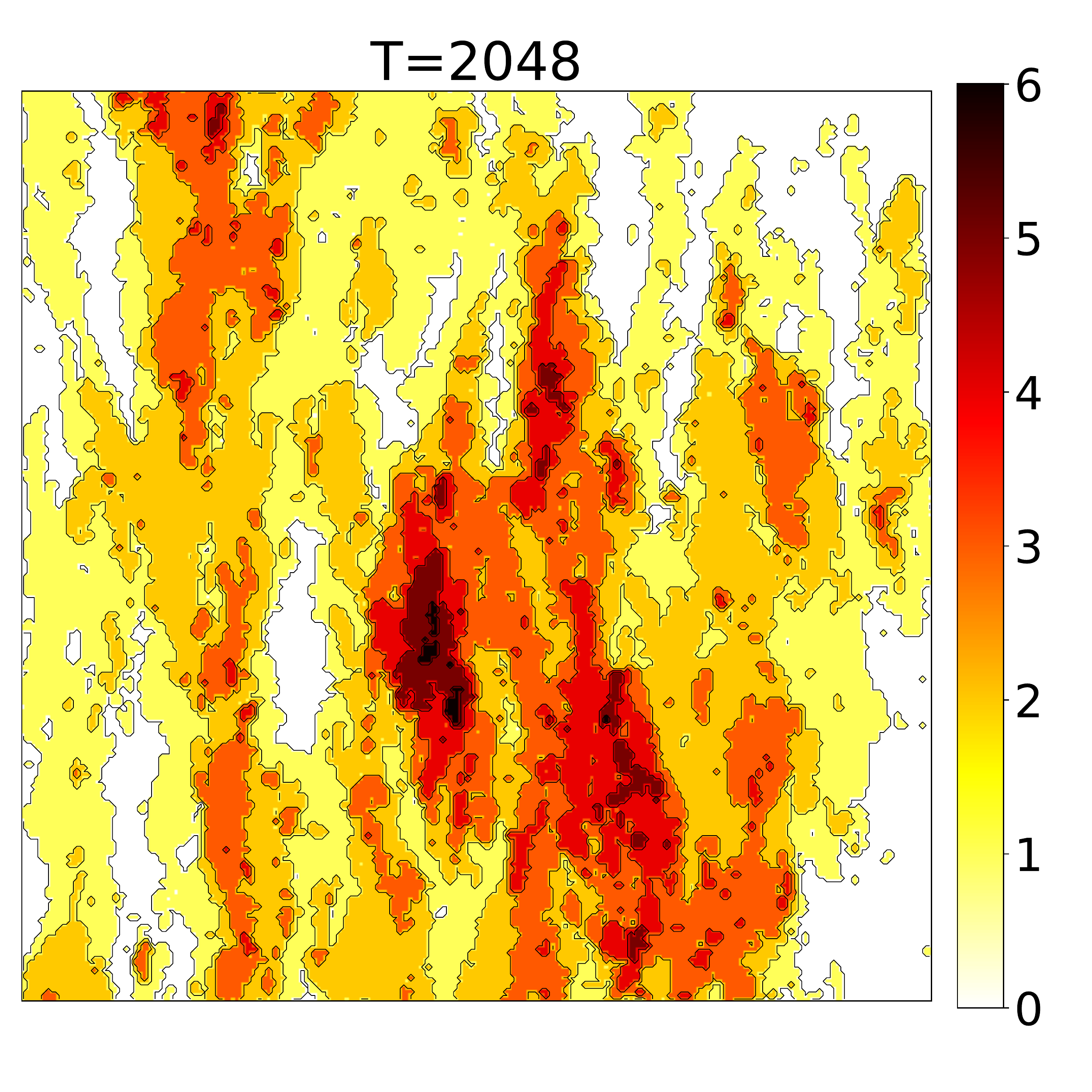}
	 	\caption{}
	 	\label{Fig:total_act_contour3}
	 \end{subfigure}
 	\begin{subfigure}[b]{0.49\linewidth}
 		\includegraphics[width=\linewidth]{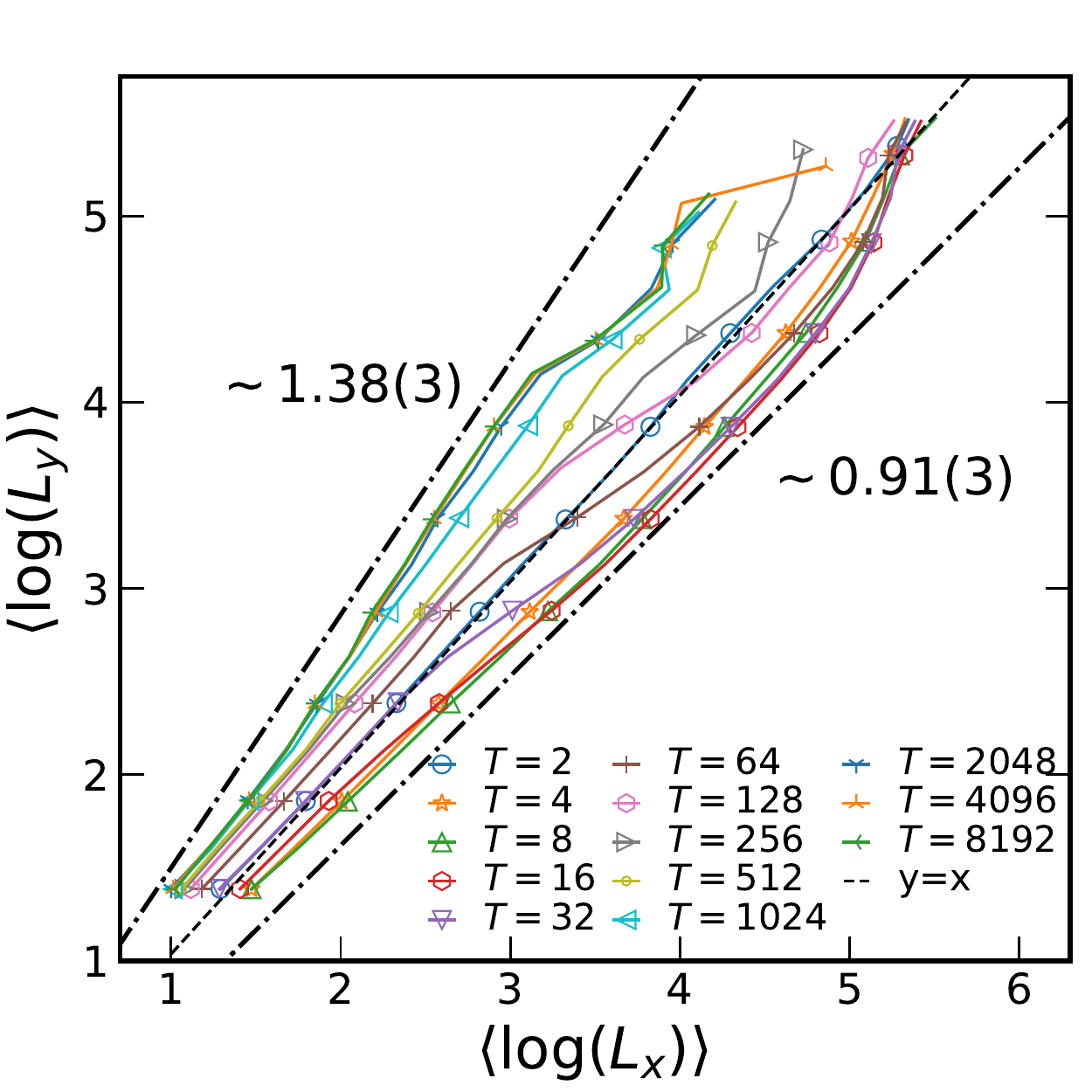}
 		\caption{}
 		\label{Fig:LxLy}
 	\end{subfigure}
 		\begin{subfigure}[b]{0.49\linewidth}
 		\includegraphics[width=\linewidth]{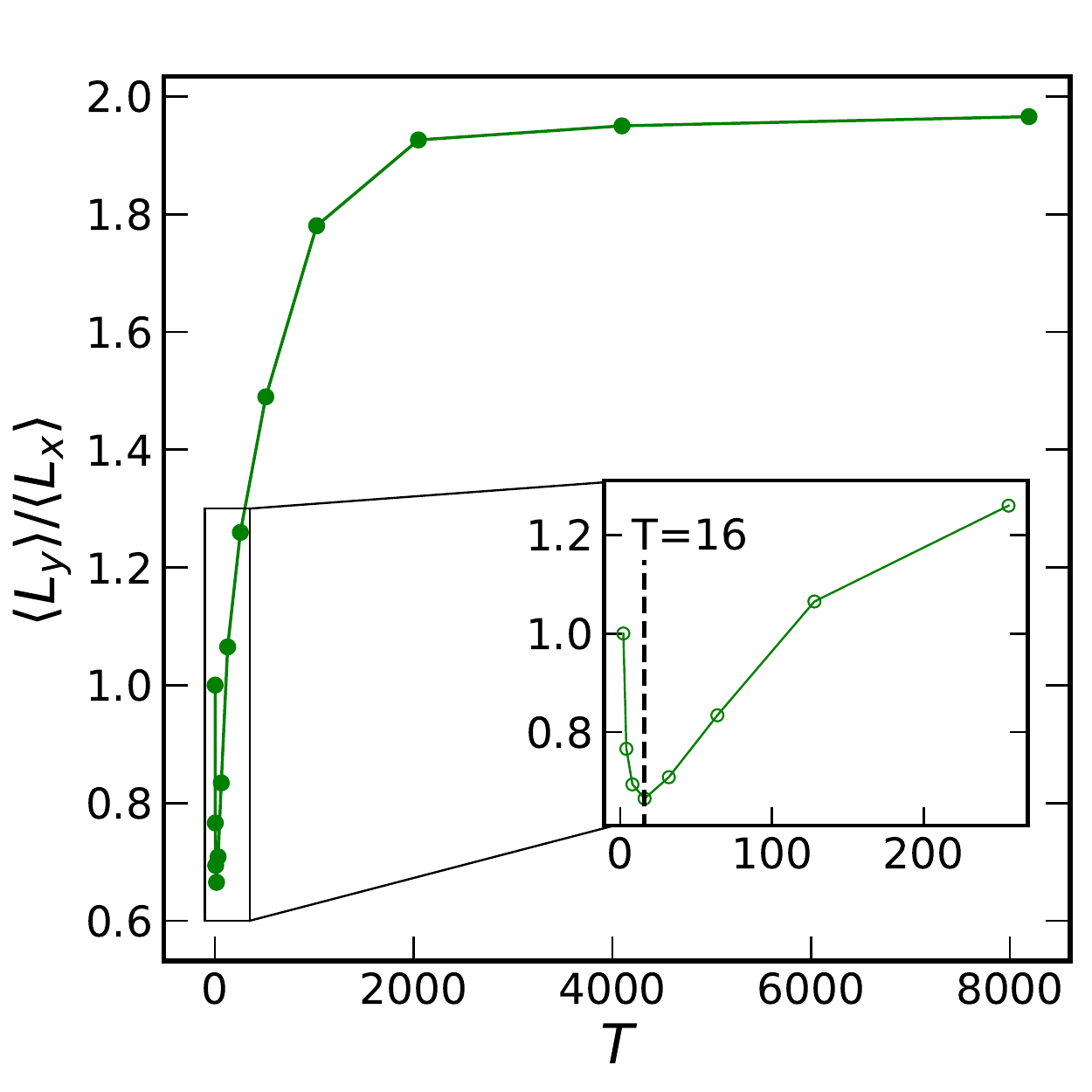}
 		\caption{}
 		\label{Fig:LxoLy}
 	\end{subfigure}
 	\centering
 	\caption{Finite activity (a) $T=2$ and $t\in[t_{\text{recurrent}}, t_{\text{recurrent}}+1000]$ time step (b) $T=64$ and $t\in[t_{\text{recurrent}}, t_{\text{recurrent}}+1000]$  time step (c) $T=2048$ and $t\in[t_{\text{recurrent}}, t_{\text{recurrent}}+10000]$  time step. For all case $L=256$. (d)  $\langle \log L_y \rangle$ in term of $\langle \log L_x \rangle$. $L_x$ and $L_y$  show a box size that covers a growing avalanche. (e)  $\frac{\langle L_y \rangle}{\langle L_x \rangle}$ in term of $T$.Snapshot time delay is $dt=100$.}
 	\label{Fig:fractalDimension}	
 \end{figure}

 To quantify this, we defined the avalanches as the chain of activities in the interval $t\in[t_{\text{recurrent}}, t_{\text{recurrent}}+1000]$, and extracted the contour lines of the resulting avalanches, see Figs.~\ref{Fig:total_act_contour1},~\ref{Fig:total_act_contour2} and~\ref{Fig:total_act_contour3} ($T=2,64$ and $2048$ respectively). We calculated the fractal dimension of the resulting avalanches by confining the avalanches in a minimal rectangle with size $L_x\times L_y$ ($L_x=x_{\text{max}}-x_{\text{min}}$, and $L_y=y_{\text{max}}-y_{\text{min}}$, where $x_{\text{max}}$ and $y_{\text{max}}$ ($x_{\text{min}}$ and $y_{\text{min}}$) are maximum (minimum) values for $x$ and $y$ respectively for the points on the external frontier of avalanches). For an isotropic avalanches one expects that $\left\langle \log L_y\right\rangle=D_f\left\langle \log L_x\right\rangle$, where $D_f=1$. We plot $\left\langle \log L_y\right\rangle $ in terms of $\left\langle \log L_x\right\rangle$ in Fig.~\ref{Fig:LxLy} and Fig.~\ref{Fig:LxoLy}. We see that there are two slopes in figure, and the graphs cross over between them. For a better understanding, let's first consider $T=2$ (the isotropic case), for which $D_f\approx 1$ as expected. Figure~\ref{Fig:LxoLy} shows that the ratio of averages $\frac{\left\langle L_y\right\rangle}{\left\langle L_x\right\rangle}$ decrease with $T$ for small $T$ values (up to $T^*\approx 16$), and then increases, and eventually saturates to $T_{\text{sat}}=1.95\pm 0.03$. A same phenomena is seen in Fig.~\ref{Fig:LxLy}, where the lowest slope is observed in $T=16$, while the highest value largest considered $T$, for which the slope is approximately $1.38(3)$. The expected orientational transition is evidenced by changing from $D_f<1$ to $D_f>1$. For intermediate $T$ values, the graphs crossover from the first to the second regime, as is seen in the figure.\\
 
\begin{figure*}
	\centering
	\includegraphics[width=0.7\linewidth]{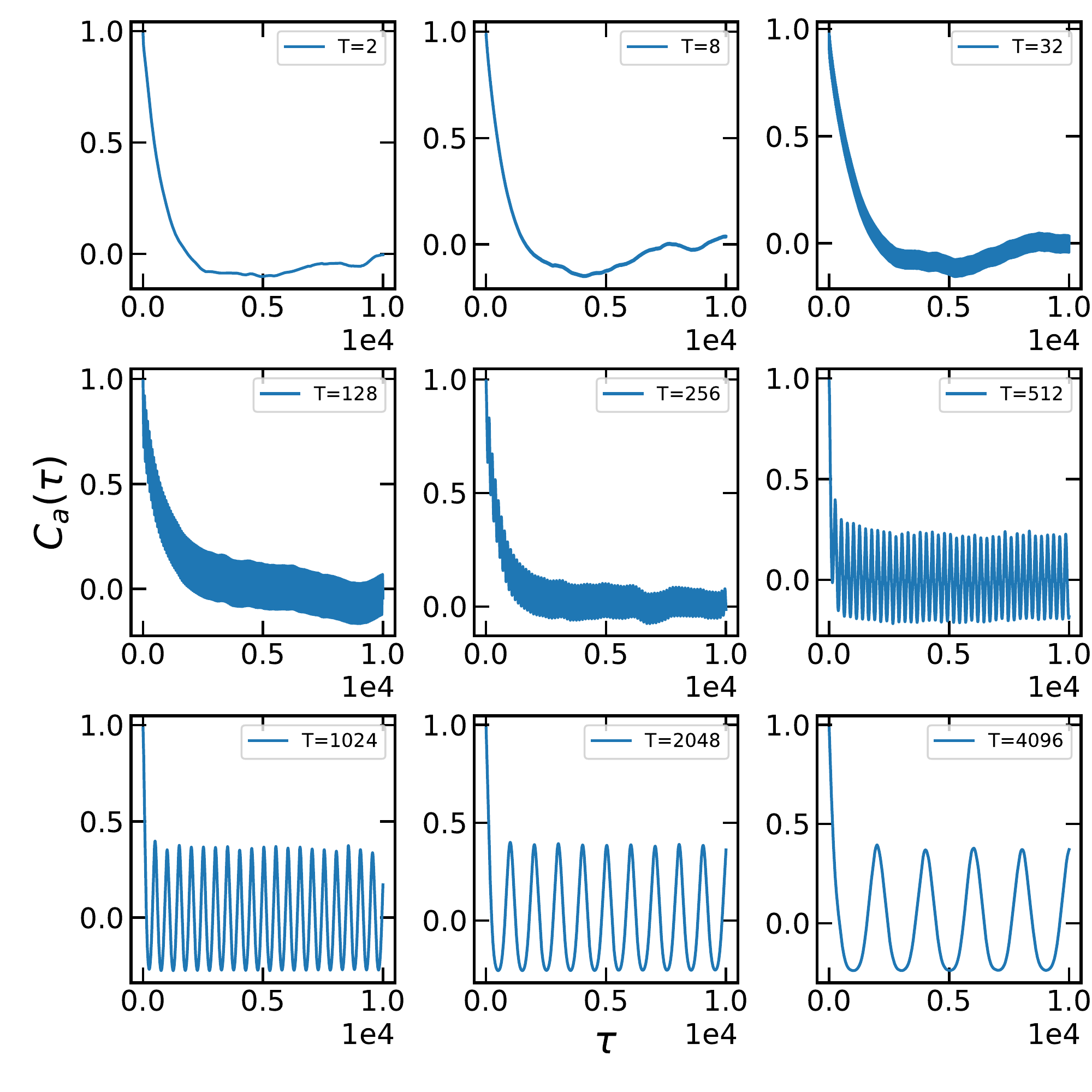}
	\caption{ The system activity ACF for various T on $L=256$ lattice size.}%
	\label{Fig:Auto_a}	
\end{figure*}

As a response to a sinusoidal external drive, the system's activity is expected to show periodic behaviour. Since the activity field is not sensitive to the sign of the external drive, one expects that the activity ACF has a half period of the one for the external drive. Fig.~\ref{Fig:Auto_a} shows this function for various rates of period. Firstly observe that the system without oscillations ($T=2$) is correlated. As $T$ increases, some tiny oscillations are born the period of which increases by increasing $T$. For example, for $T=4096$, the period of ACF is $\tau_0=2048=T/2$. To visualize the structure of the oscillations and the characteristic periods more explicitly, one should study the power spectrum of the time series.     
\begin{figure*}
	\begin{subfigure}[b]{0.3\linewidth}
		\includegraphics[width=\linewidth]{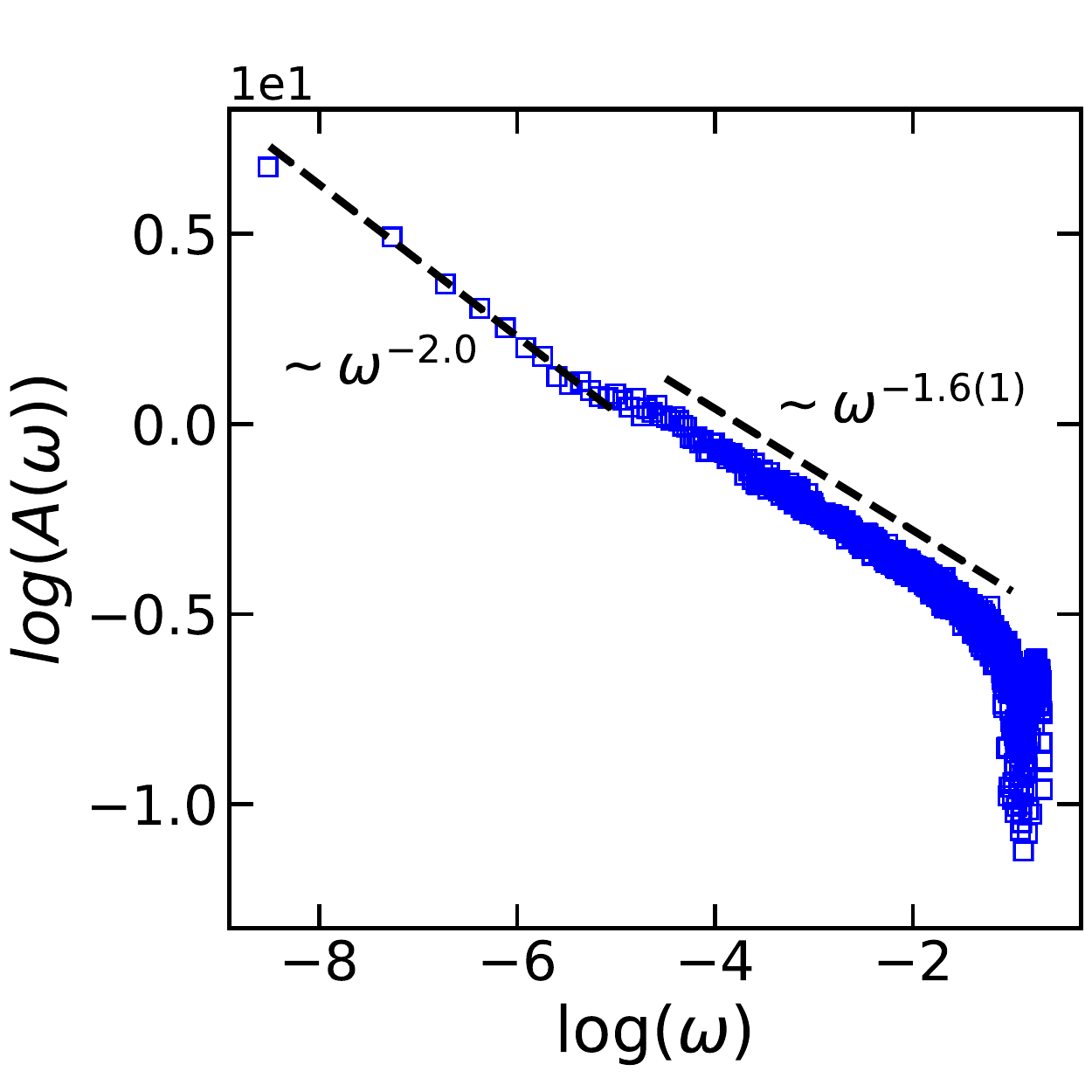}
		\caption{$T=2$}
		\label{T=2_1}
	\end{subfigure}
	\begin{subfigure}[b]{0.3\linewidth}
		\includegraphics[width=\linewidth]{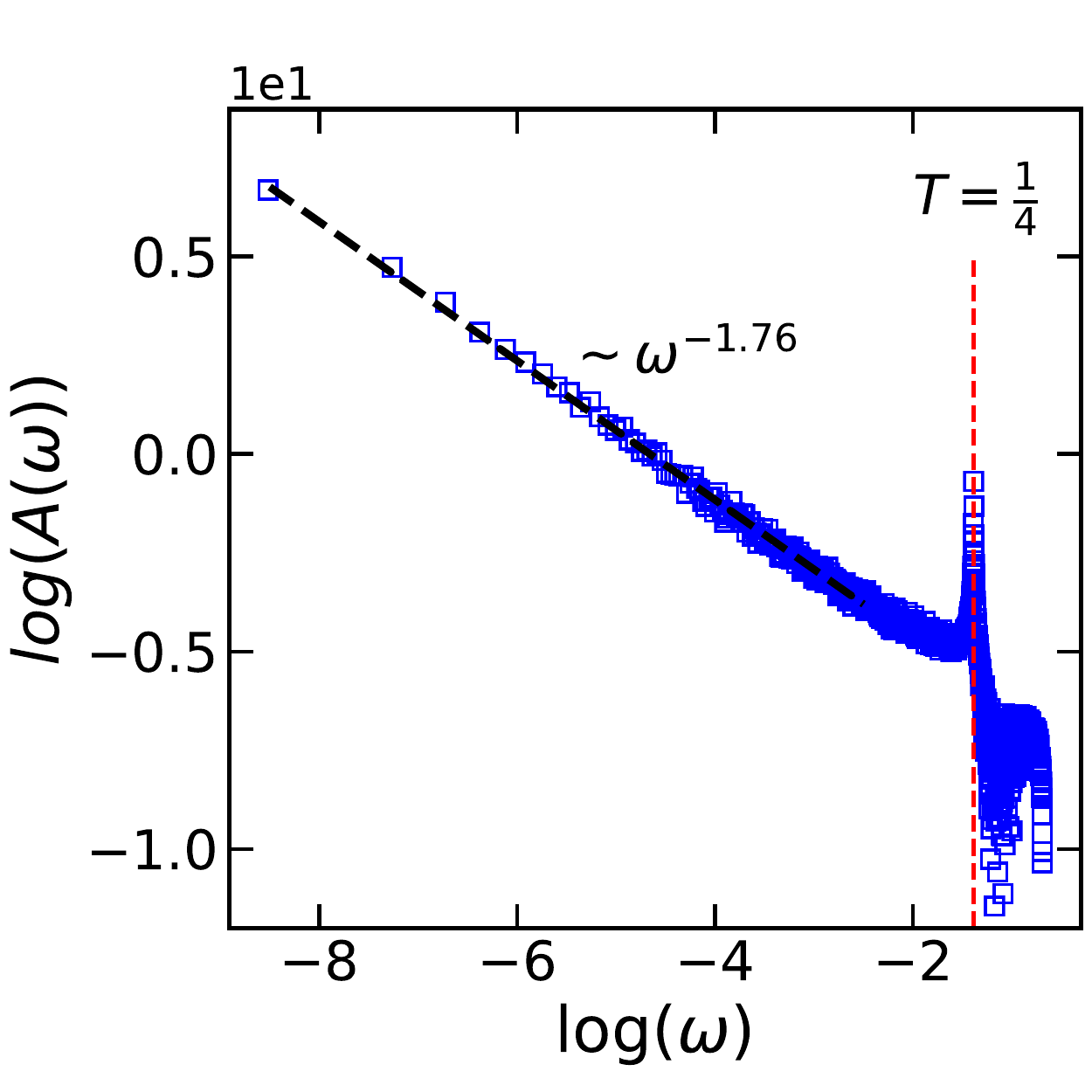}
		\caption{$T=8$}
		\label{T=8_2}
	\end{subfigure}
	\begin{subfigure}[b]{0.3\linewidth}
		\includegraphics[width=\linewidth]{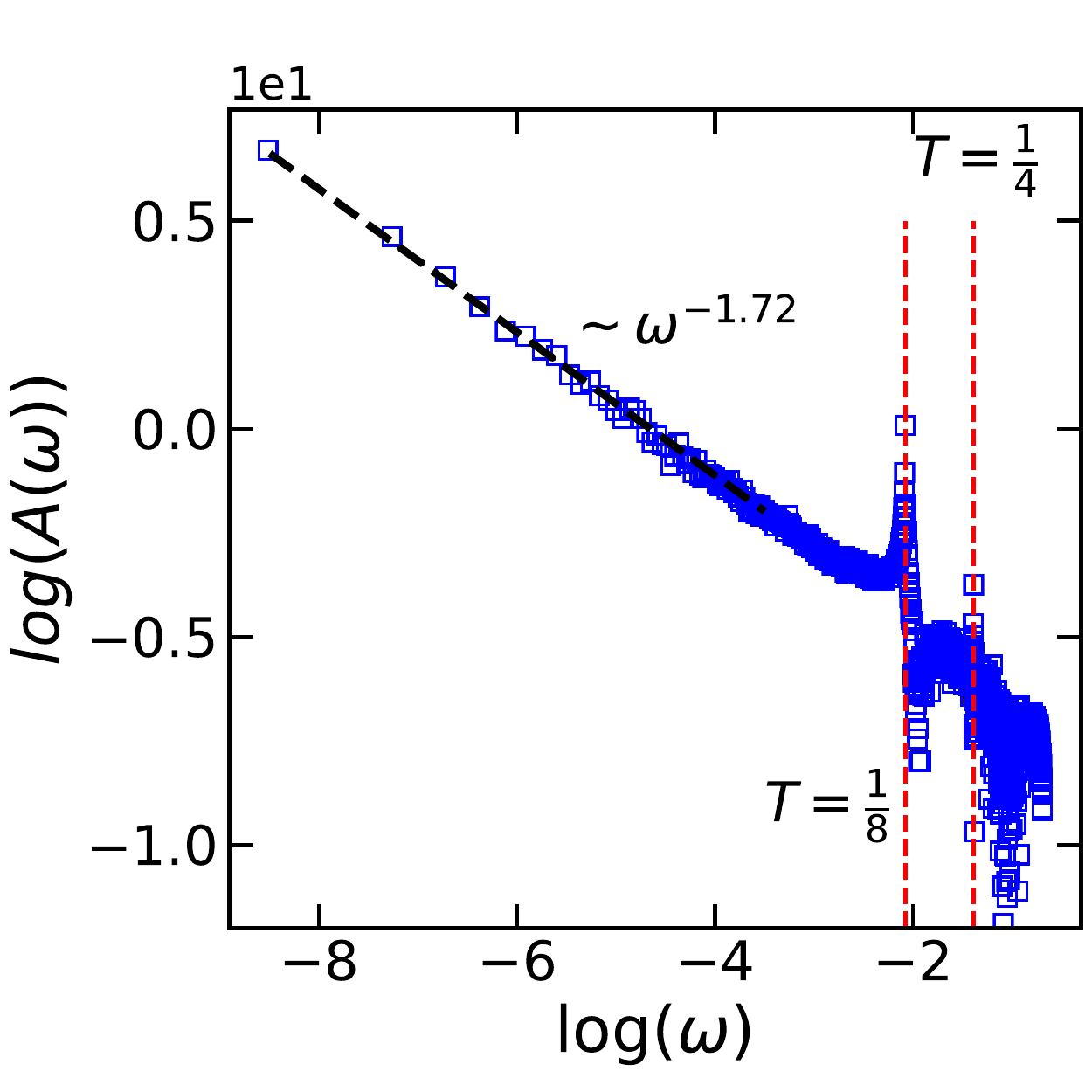}
		\caption{$T=16$}
		\label{T=16_2}
	\end{subfigure}
	\begin{subfigure}[b]{0.3\linewidth}
		\includegraphics[width=\linewidth]{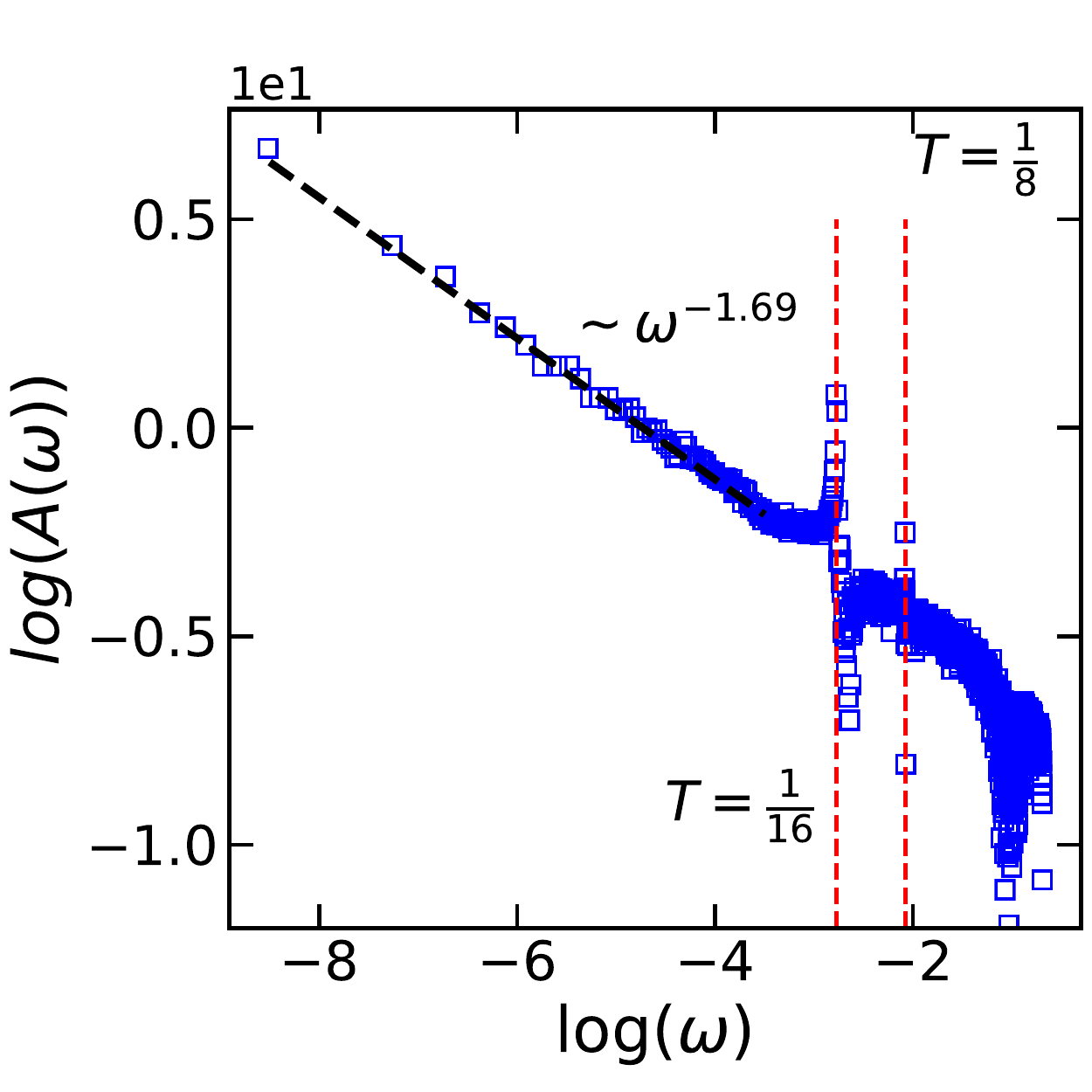}
		\caption{$T=32$}
		\label{T=32_2}
	\end{subfigure}
	\begin{subfigure}[b]{0.3\linewidth}
		\includegraphics[width=\linewidth]{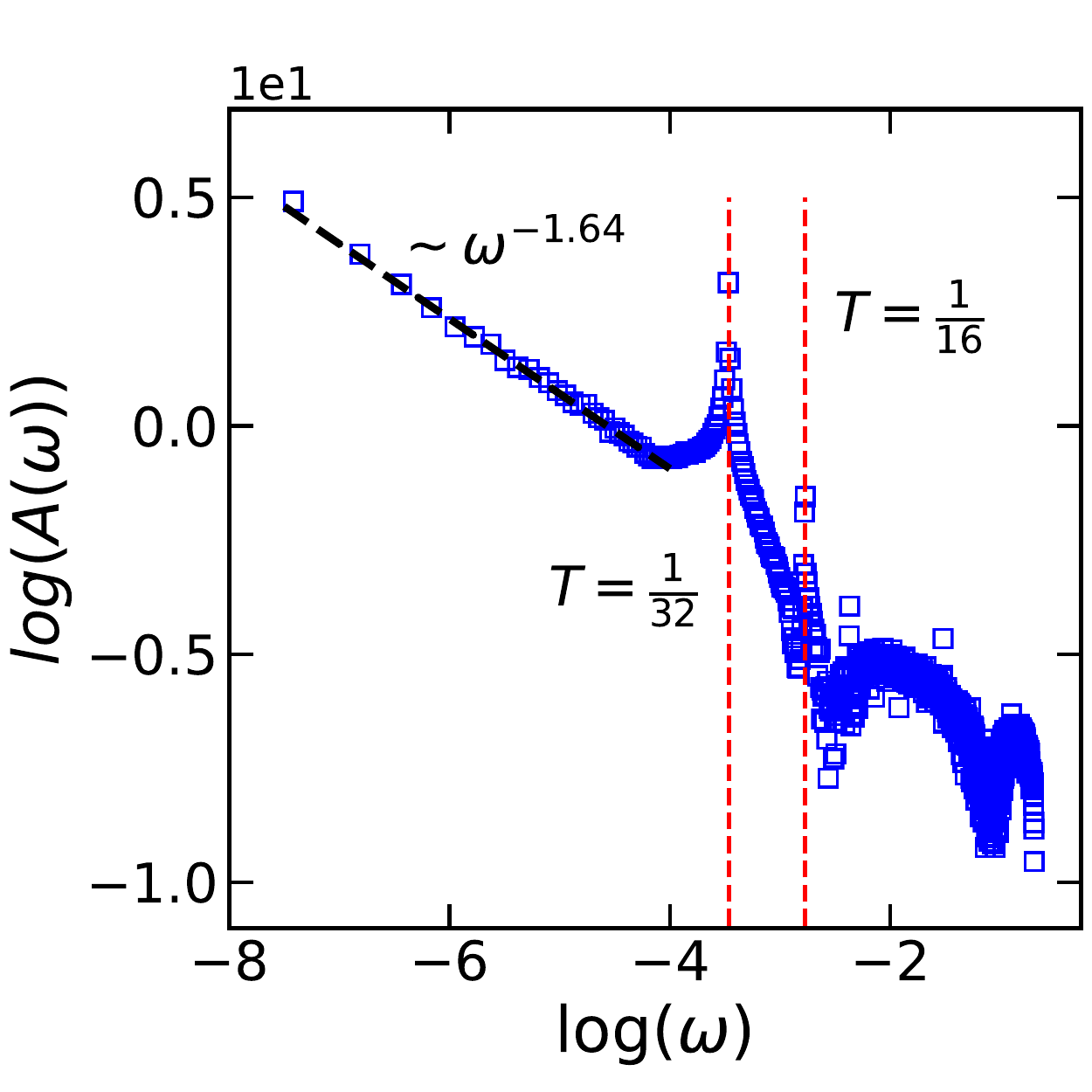}
		\caption{$T=64$}
		\label{T=64_2}
	\end{subfigure}
	\begin{subfigure}[b]{0.3\linewidth}
		\includegraphics[width=\linewidth]{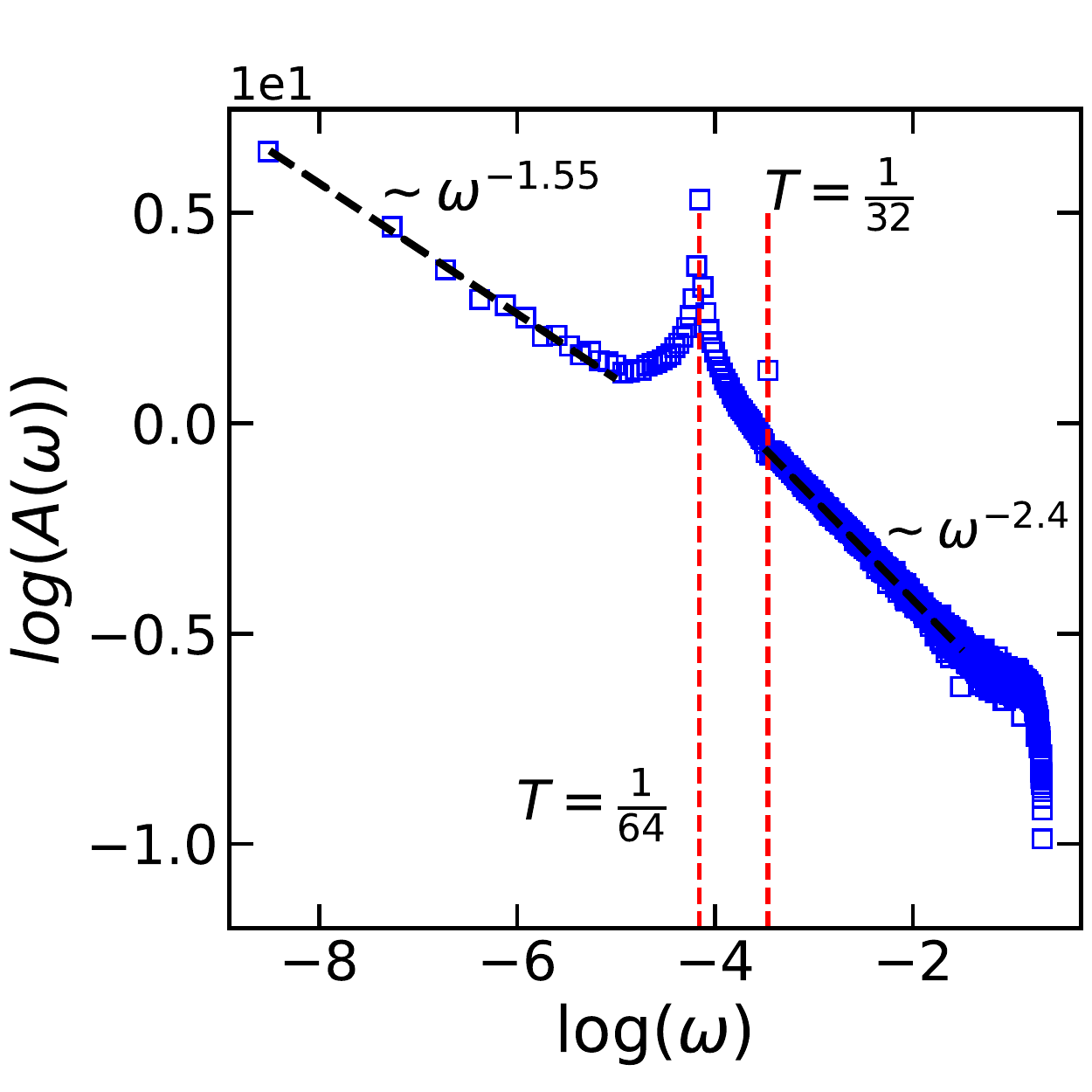}
		\caption{$T=128$}
		\label{T=128_2}
	\end{subfigure}
	\begin{subfigure}[b]{0.3\linewidth}
		\includegraphics[width=\linewidth]{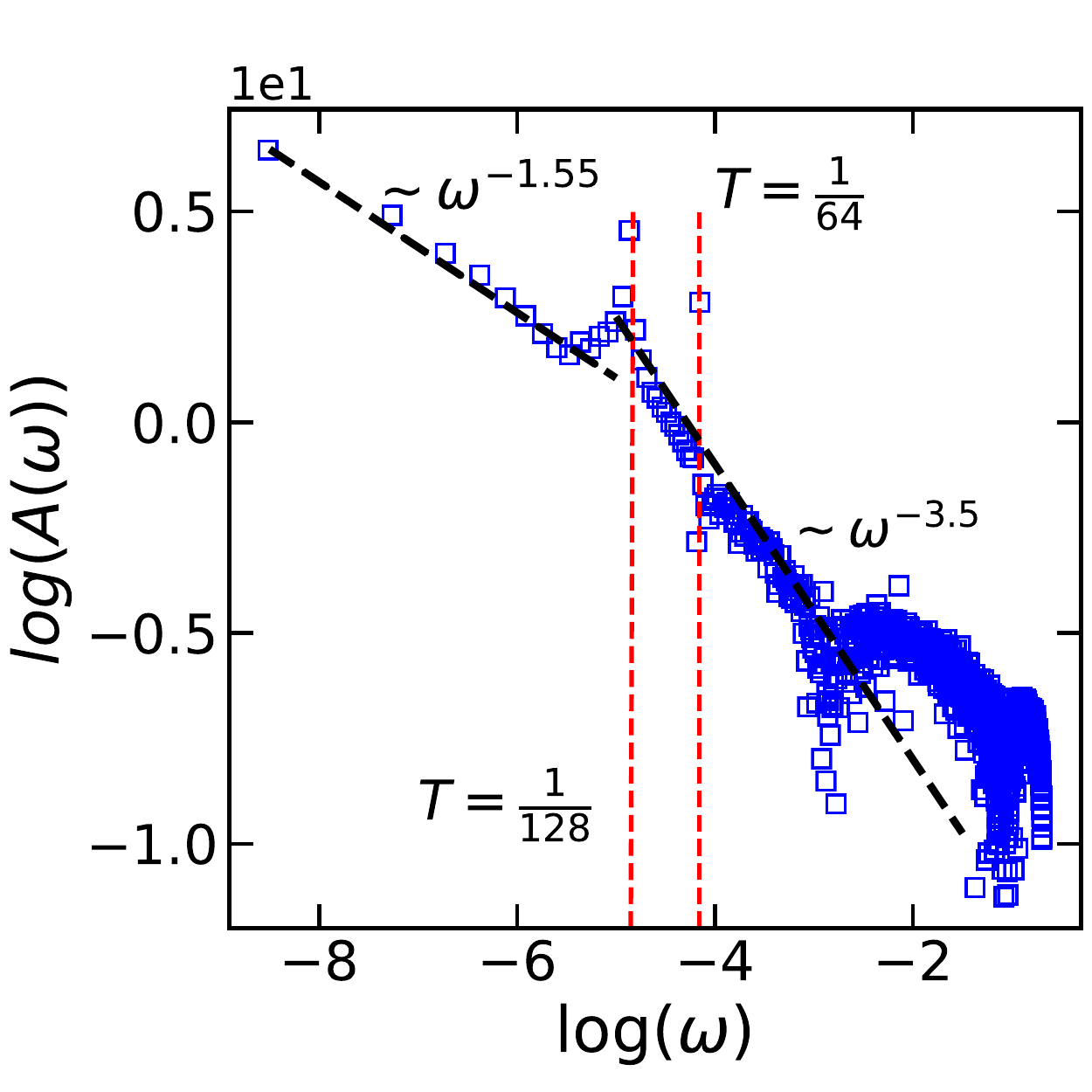}
		\caption{$T=256$}
		\label{T=256_2}
	\end{subfigure}
	\begin{subfigure}[b]{0.3\linewidth}
		\includegraphics[width=\linewidth]{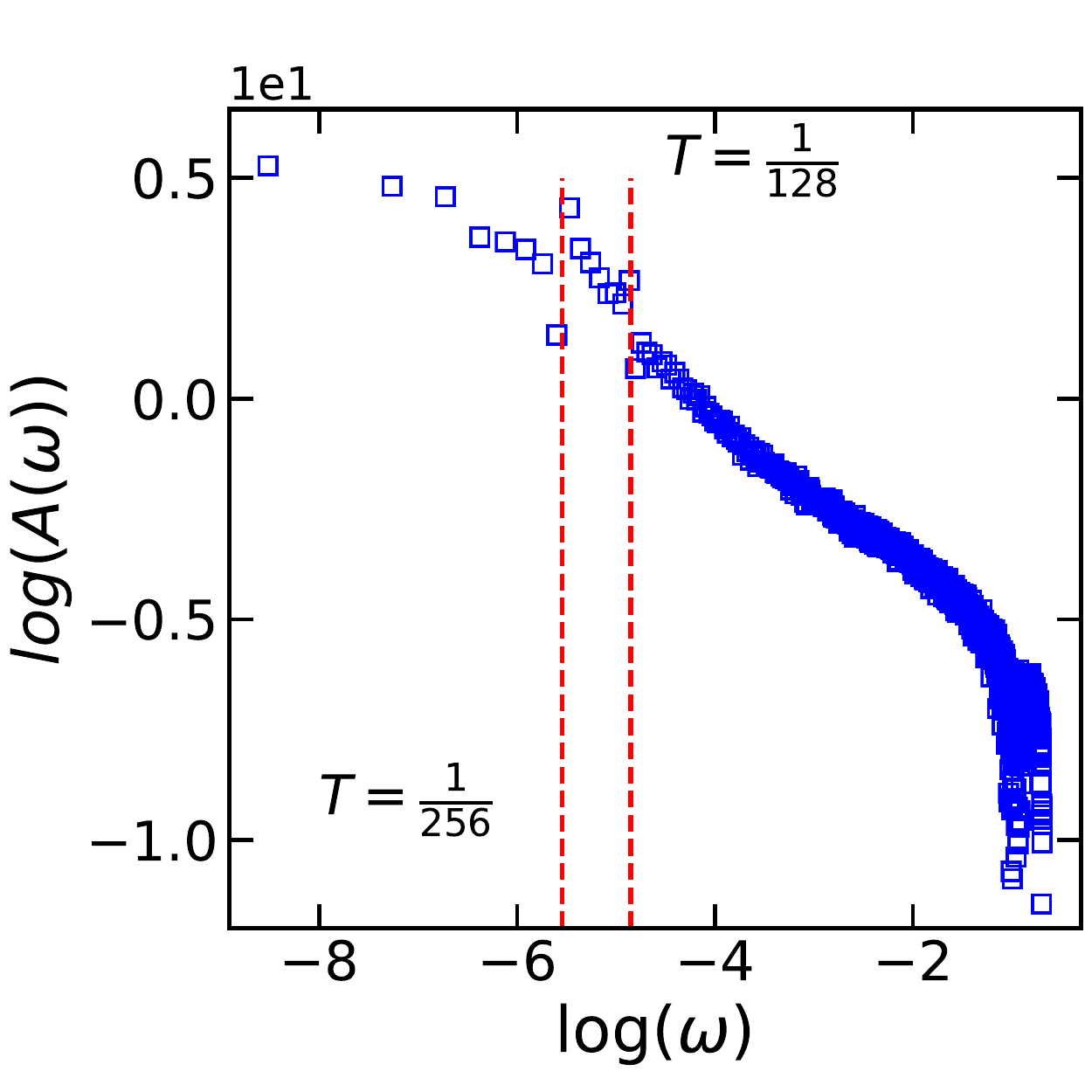}
		\caption{$T=512$}
		\label{T=512_2}
	\end{subfigure}
	\begin{subfigure}[b]{0.3\linewidth}
		\includegraphics[width=\linewidth]{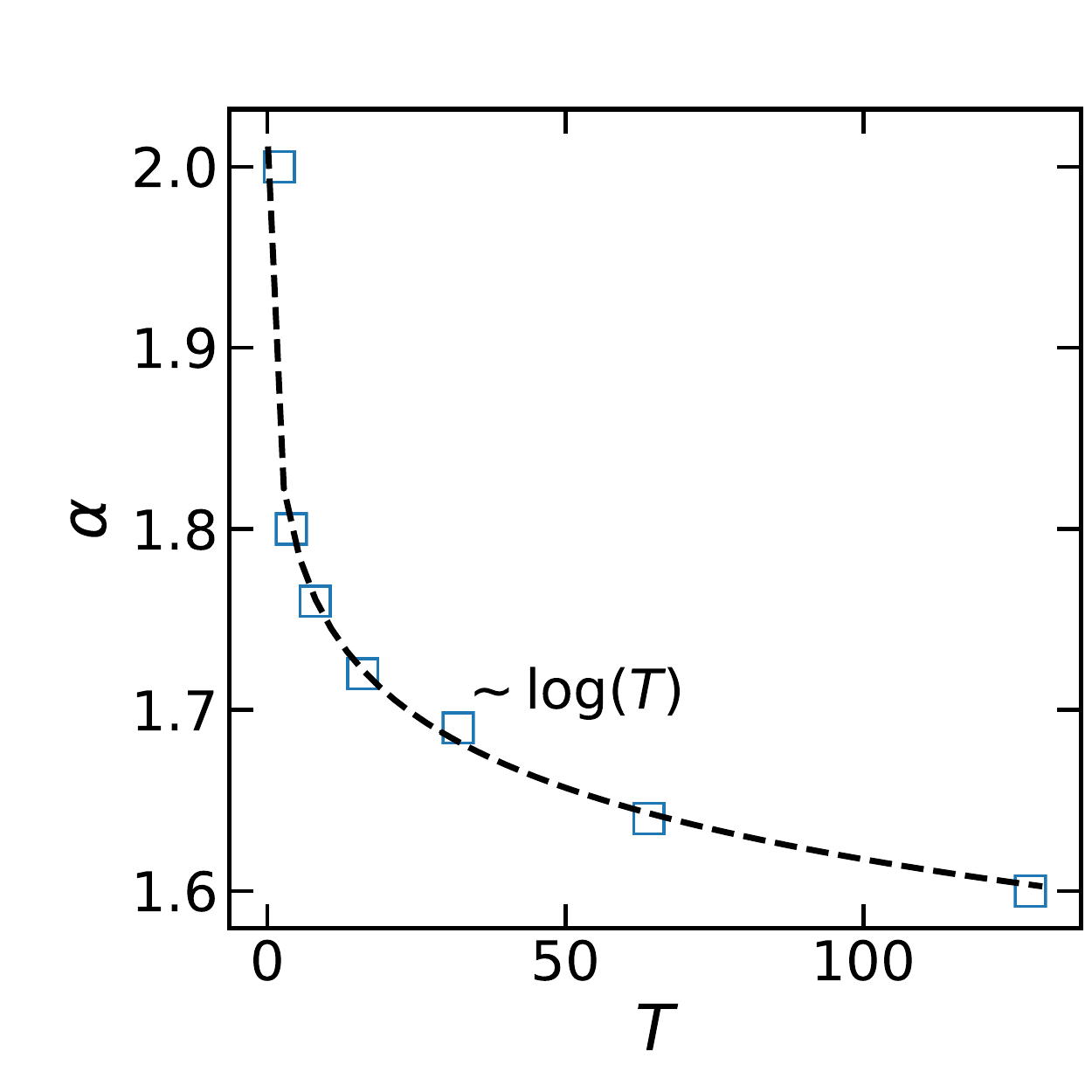}
		\caption{}
		\label{Fig:alpha-PS}
	\end{subfigure}
	\centering
	\caption{The system activity PSD for various $T$ at $L = 256$ lattice size. (i) $\alpha_{\text{PS}}^{(1)}$ exponents in term of $T$ for the PSD of the activity time series (i.e., $A(\omega) \sim f^{-\alpha}$). The fit is logarithm of $T$.}
	\label{PSD_a}	
\end{figure*}

PSDs are shown in Fig.~\ref{PSD_a}, where a multi-fractal structure (self-similar) structure is observed, reflected in $A(\omega)\propto \omega^{\alpha_{\text{PS}}}$, where $\alpha_{\text{PS}}$ depends on the scale. Observe how the non-oscillating system ($T=2$) exhibits two regimes: one with the exponent $\alpha_{\text{PS}}\simeq 2$, and another with $\alpha_{\text{PS}}\simeq 1.6$. This has already been known that avalanches in sandpiles are multi-fractals, but since ACFs (for the BTW model) exponentially decay with time, $\alpha_{\text{PS}}=2$ (red noise) is expected for them as numerically has been confirmed. Therefore the exponents of the OSM are new and have not been reported before. Additionally, PSD functions show peaked structures, having their roots in the fact that there are driven oscillations in the ACFs as we just saw. Indeed, for $\omega\gg \frac{2\pi}{T}$ it has information about single avalanches, while for $\omega\ll \frac{2\pi}{T}$ it tells about long-time correlations of activity. The regimes are identified with respect to two pronounced peaks which have been identified in the figures, one of which is at $f^{-1} =T/2$ and the other one is $f^{-1} =T/4$. One can study the long-time correlations more effectively by subtracting the power spectrum at frequencies $\omega\gg \frac{2\pi}{T}$. The power spectrum exponent in the small frequency regime ($\alpha_{\text{PS}}^{(1)}$) decreases by increasing $T$. This is shown in Fig.~\ref{Fig:alpha-PS}, where a logarithmic decay $\alpha_{\text{PS}}^{(1)}=a-b\log T$ is observed, where $a=1.88\pm0.04$ and $b=0.06\pm 0.01$. This fitting is not valid for all ranges of $T$, i.e. the best fitting is for the range $T\lesssim 128$, where. For a self-similar mono-fractal time series with a Hurst exponent $H$, the power spectrum is expected to behave like $\omega^{-2H+1}$~\cite{gharari2021space,samorodnitsky2016stochastic}, which corresponds in OSM to the Hurst exponent $H=\frac{1}{2}\left(1+a-b\log T \right) $. While avalanches in OSM are multi-fractal, a more comprehensive multi-fractal analysis is necessary to determine the spectrum of the Hurst exponent. \\


\begin{figure*}
	\begin{subfigure}[b]{0.3\linewidth}
		\includegraphics[width=\linewidth]{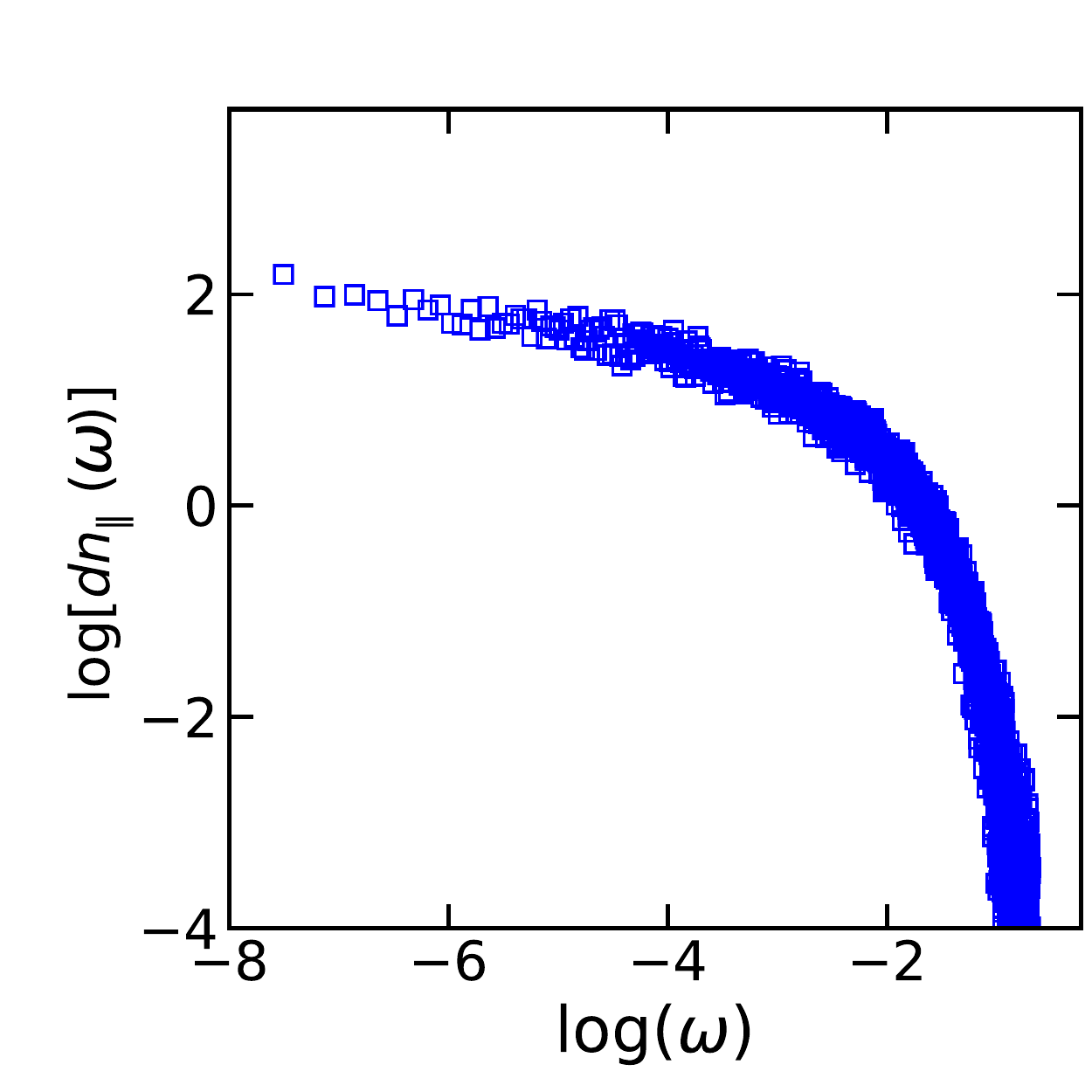}
		\caption{$T=2$}
		\label{T=2_6}
	\end{subfigure}
	\begin{subfigure}[b]{0.3\linewidth}
		\includegraphics[width=\linewidth]{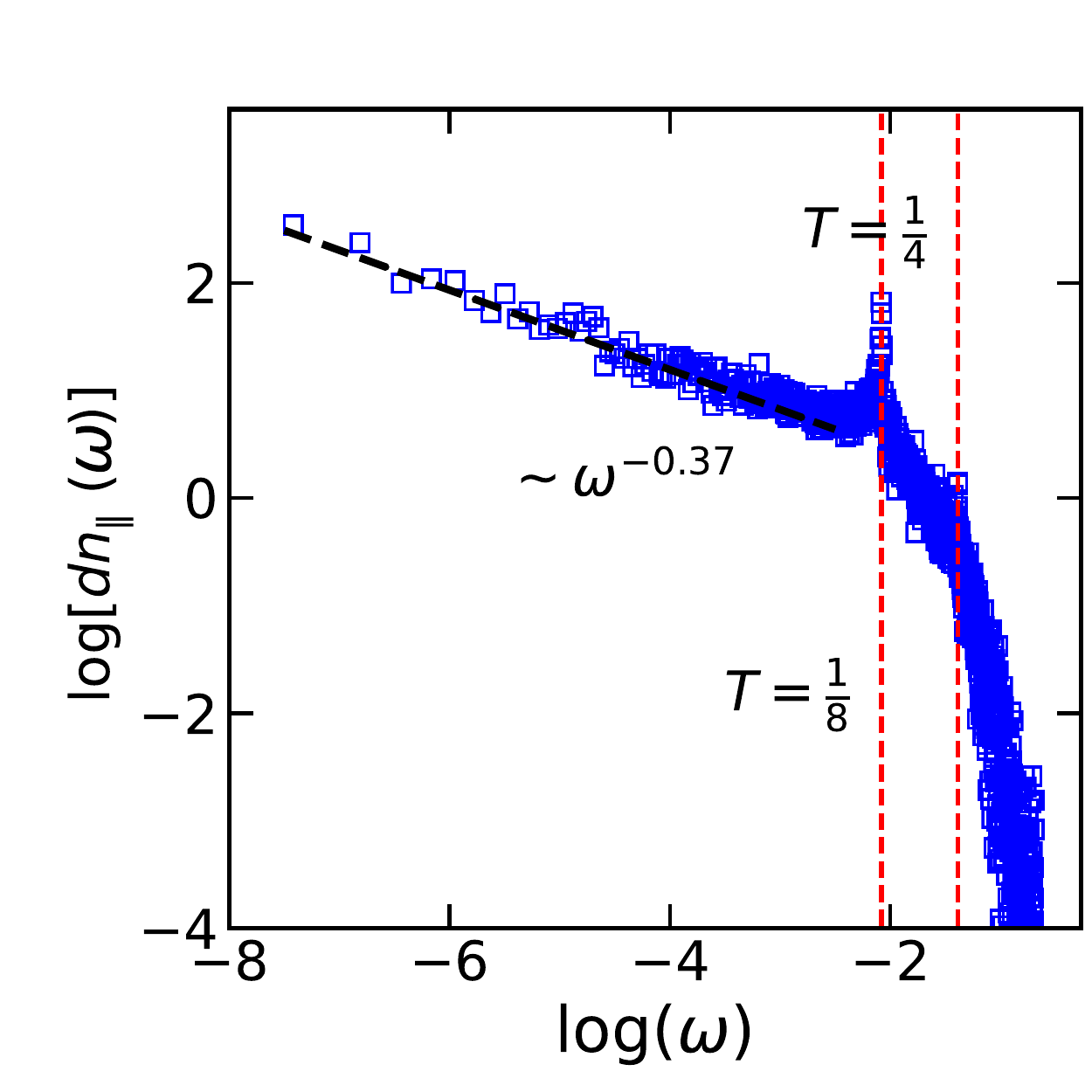}
		\caption{$T=8$}
		\label{T=8_6}
	\end{subfigure}
	\begin{subfigure}[b]{0.3\linewidth}
		\includegraphics[width=\linewidth]{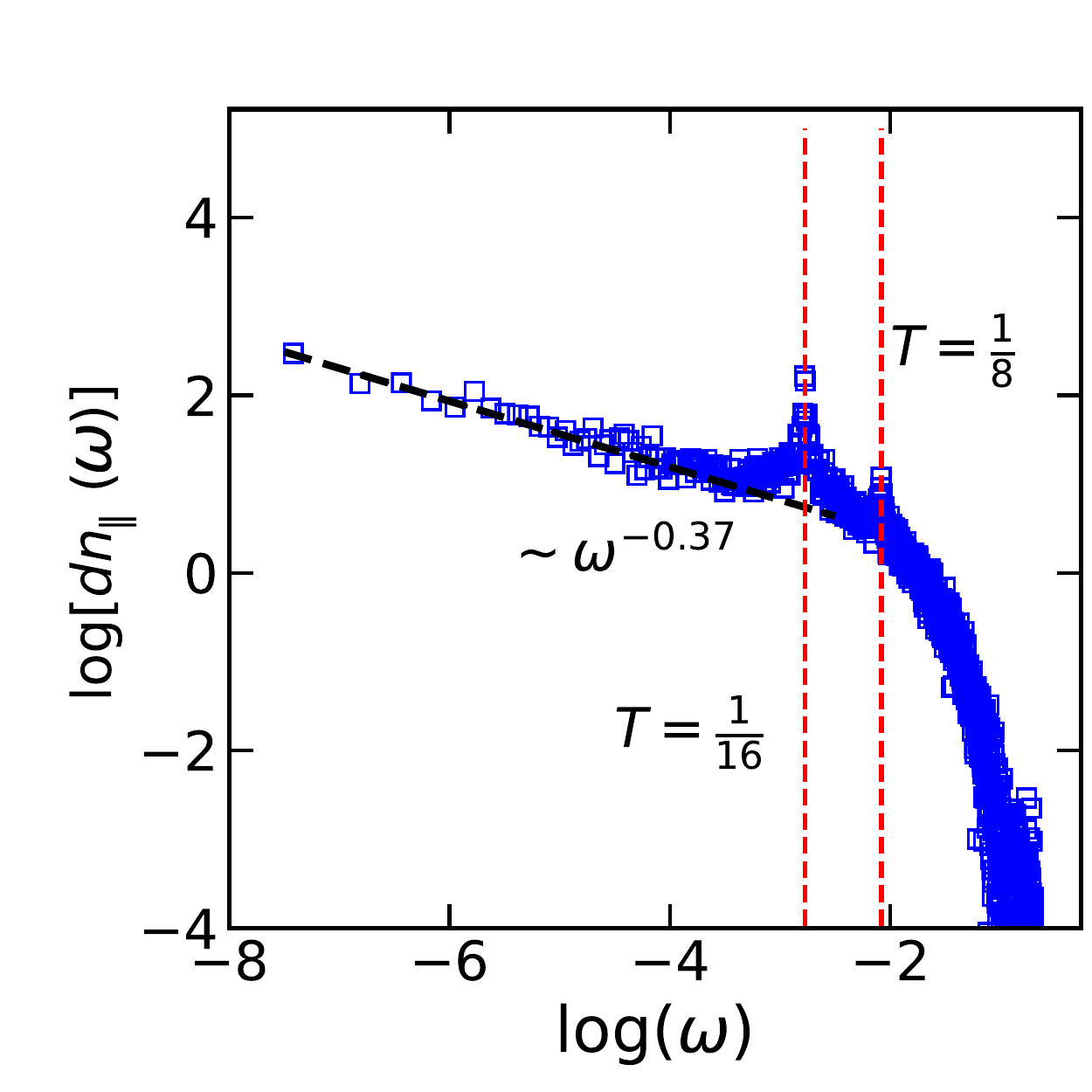}
		\caption{$T=16$}
		\label{T=16_6}
	\end{subfigure}
	\begin{subfigure}[b]{0.3\linewidth}
		\includegraphics[width=\linewidth]{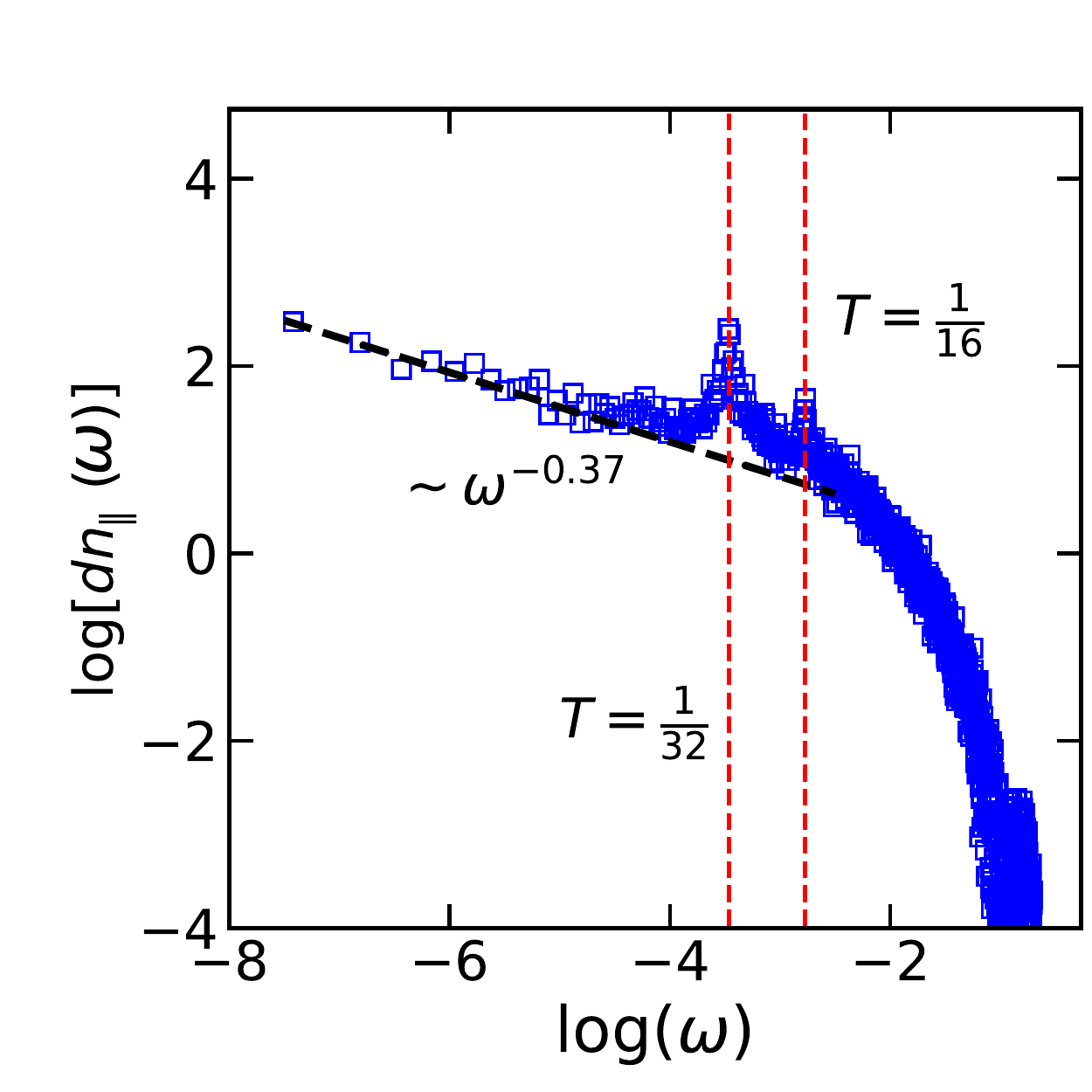}
		\caption{$T=32$}
		\label{T=32_6}
	\end{subfigure}
	\begin{subfigure}[b]{0.3\linewidth}
		\includegraphics[width=\linewidth]{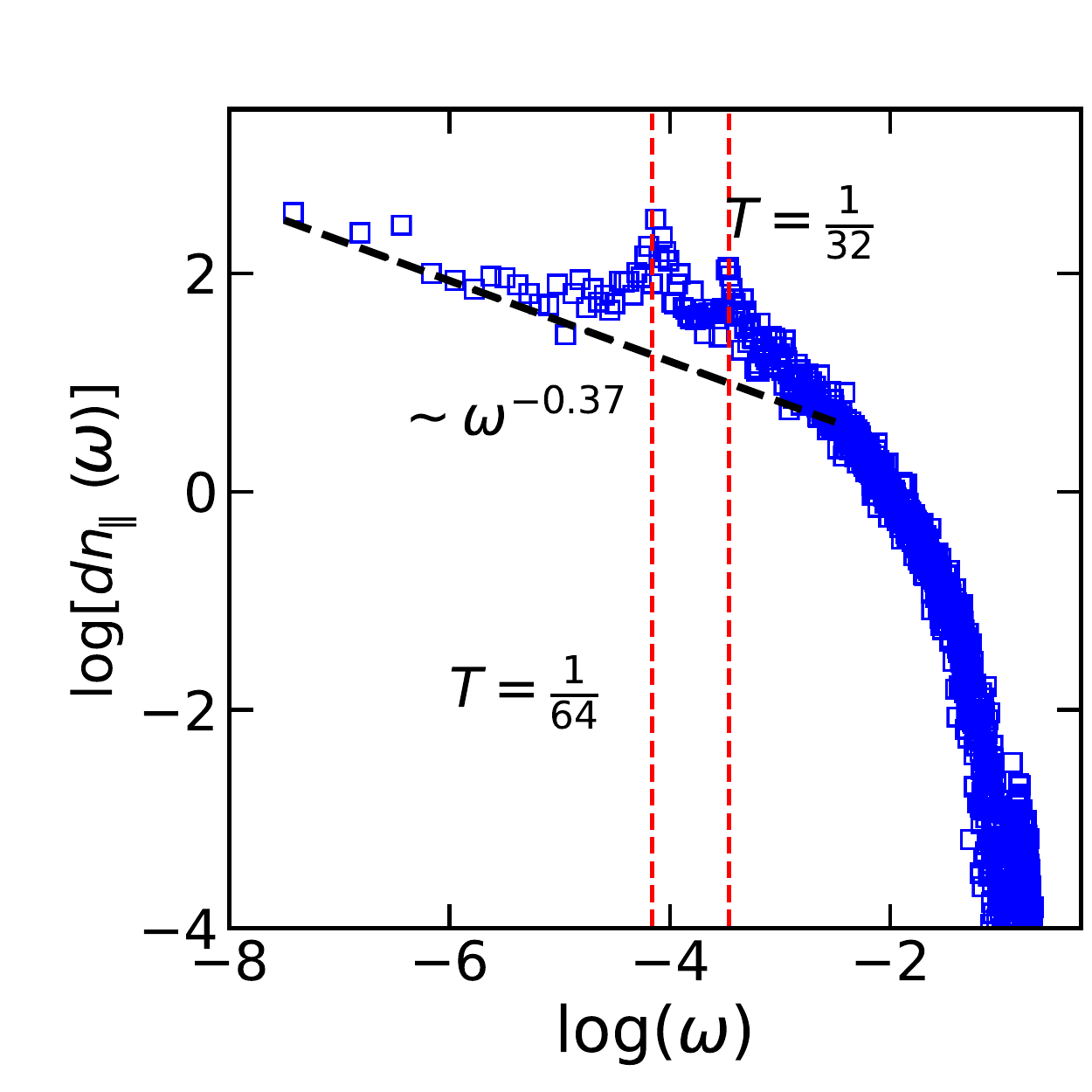}
		\caption{$T=64$}
		\label{T=64_6}
	\end{subfigure}
	\centering
	\caption{PSD for sand dissipation in the parallel direction for various $T$ at $L = 256$ lattice size.}
	\label{Fig:PSD_parallel}	
\end{figure*}

\begin{figure*}
	\begin{subfigure}[b]{0.3\linewidth}
		\includegraphics[width=\linewidth]{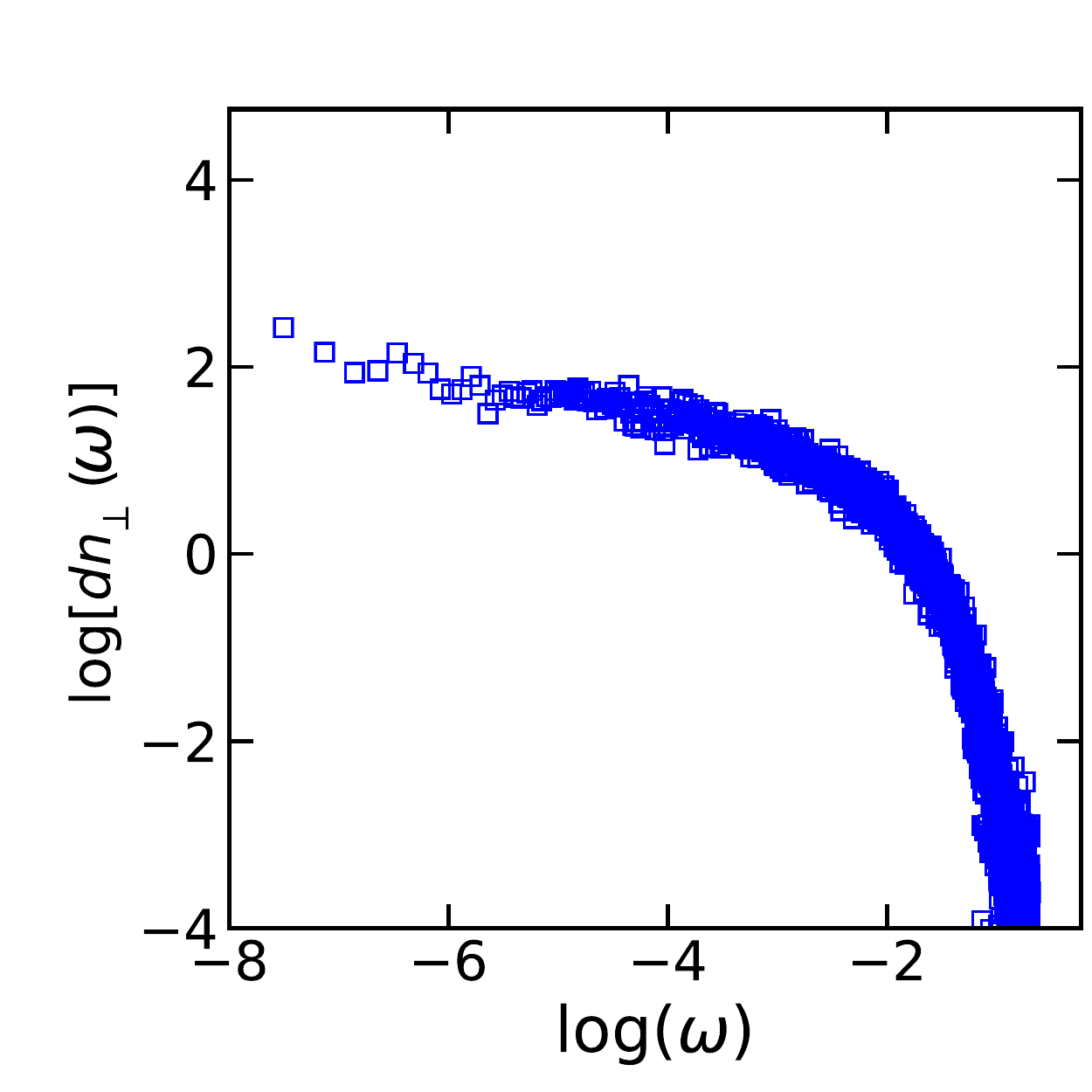}
		\caption{$T=2$}
		\label{T=2_4}
	\end{subfigure}
	\begin{subfigure}[b]{0.3\linewidth}
		\includegraphics[width=\linewidth]{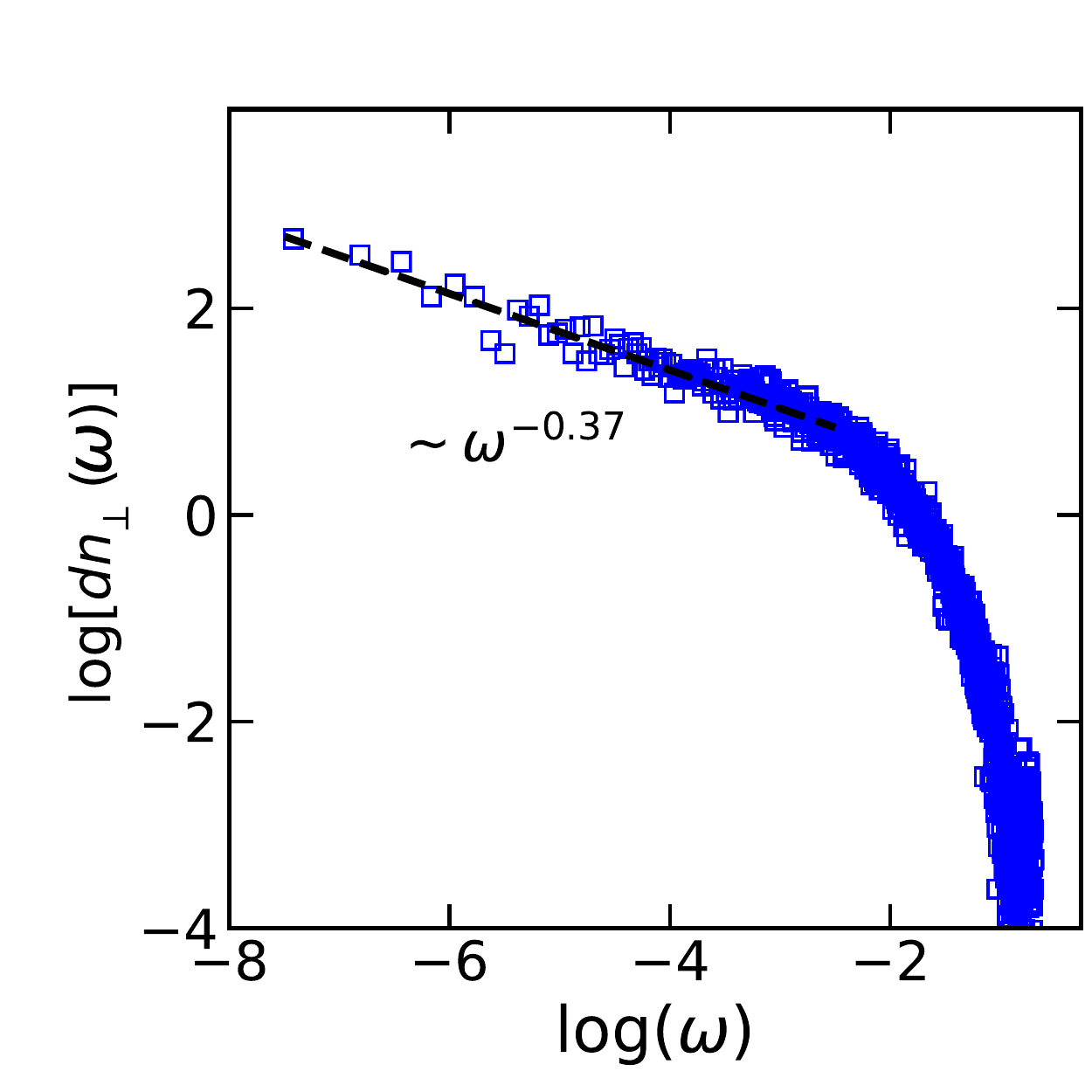}
		\caption{$T=8$}
		\label{T=8_4}
	\end{subfigure}
	\begin{subfigure}[b]{0.3\linewidth}
		\includegraphics[width=\linewidth]{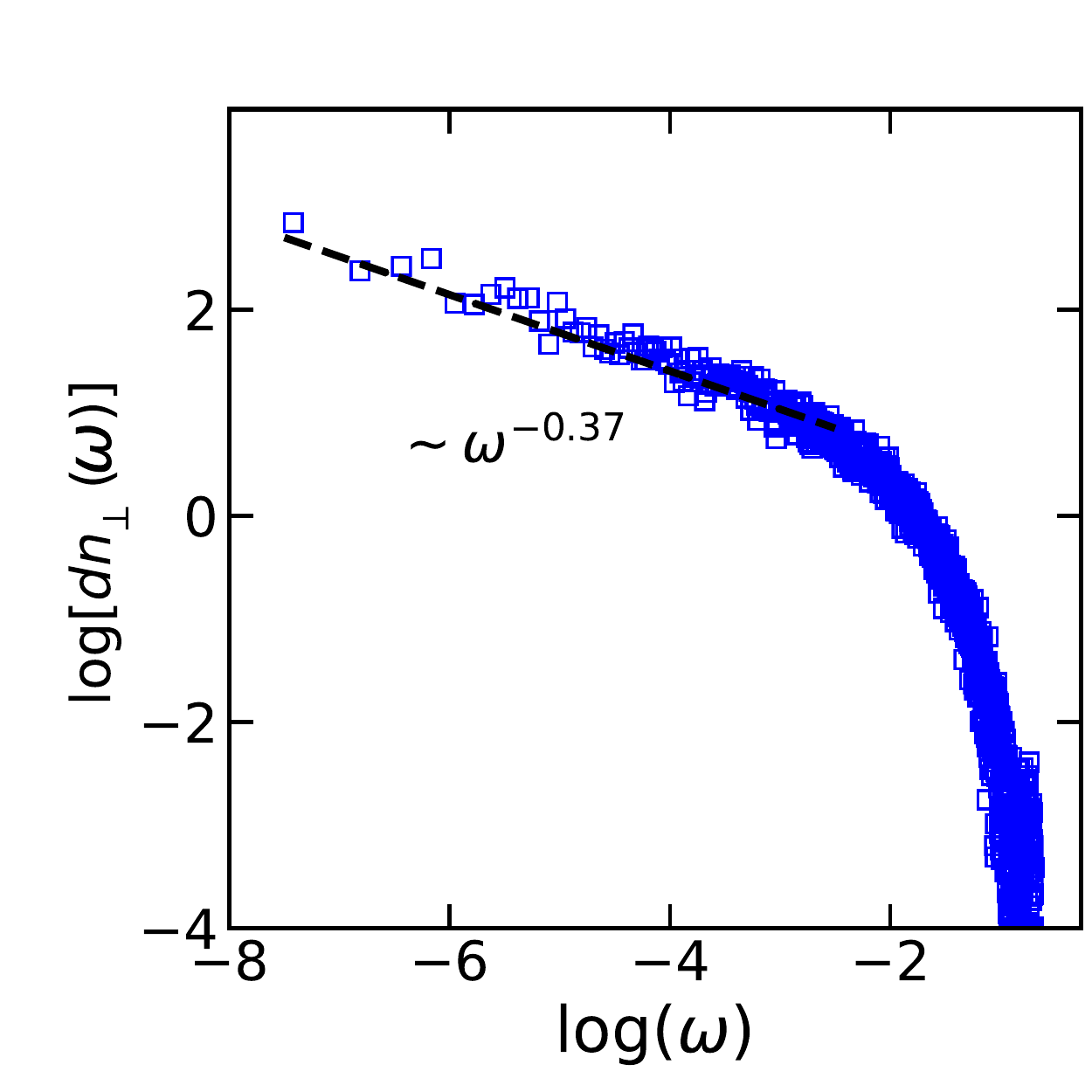}
		\caption{$T=16$}
		\label{T=16_4}
	\end{subfigure}
	\begin{subfigure}[b]{0.3\linewidth}
		\includegraphics[width=\linewidth]{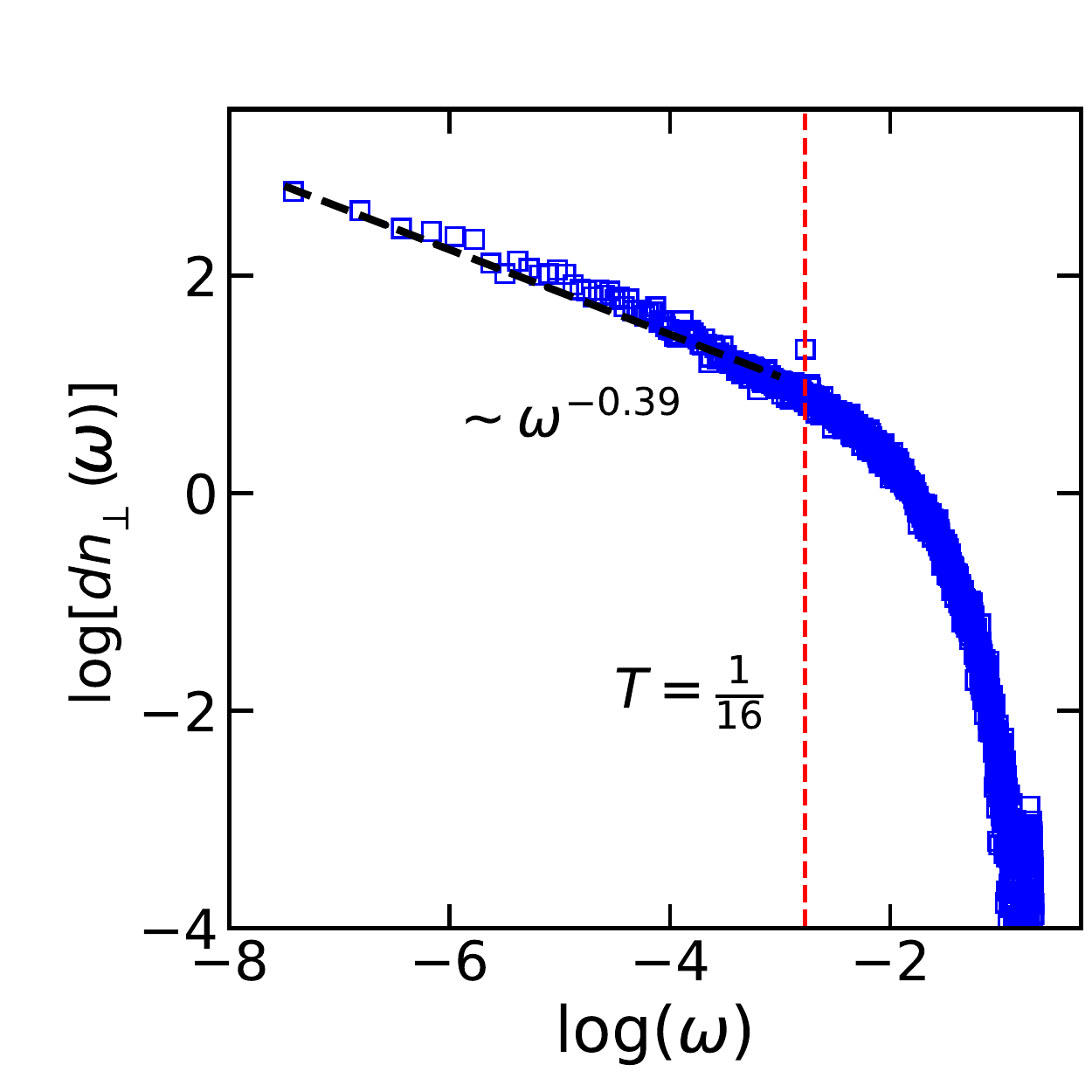}
		\caption{$T=32$}
		\label{T=32_4}
	\end{subfigure}
	\begin{subfigure}[b]{0.3\linewidth}
		\includegraphics[width=\linewidth]{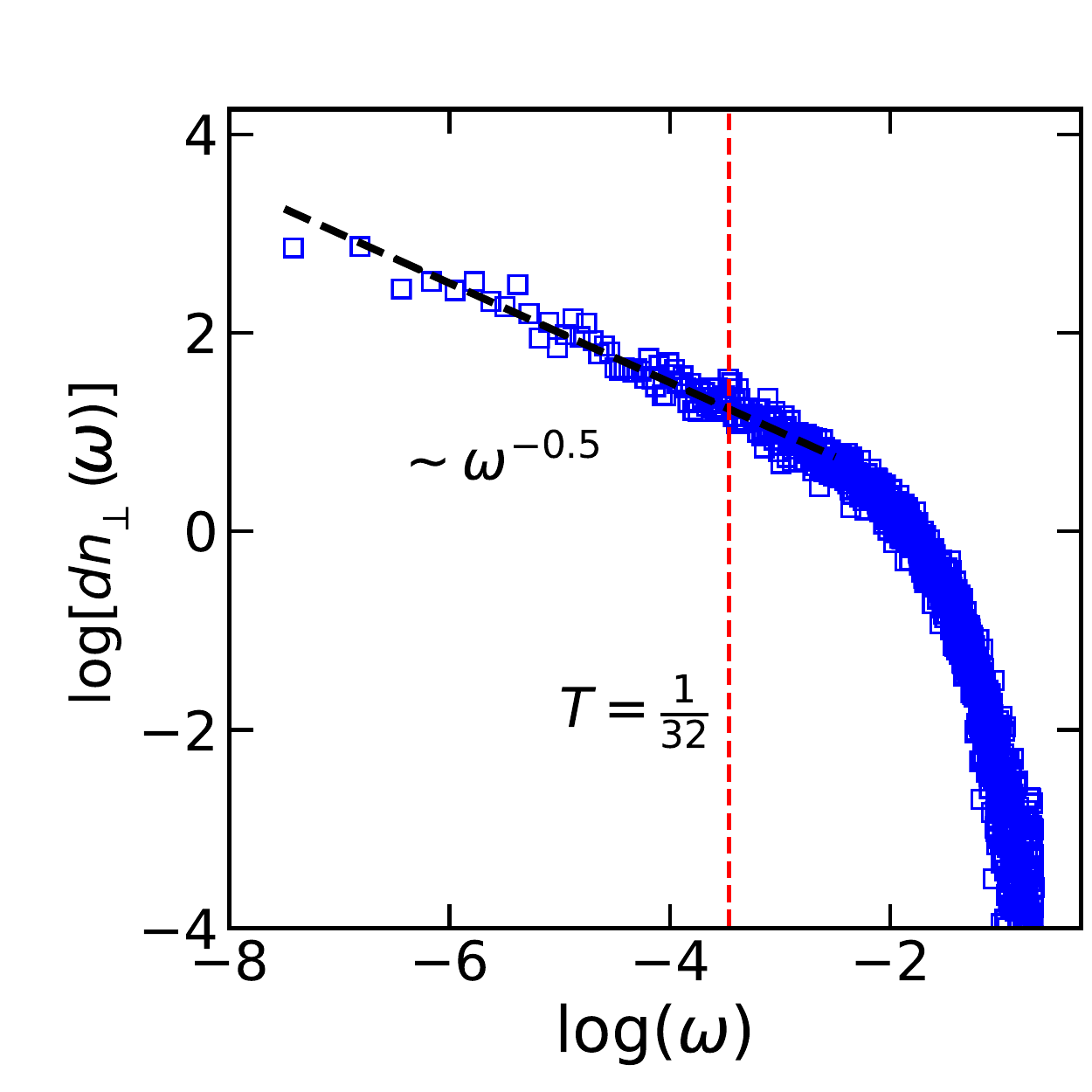}
		\caption{$T=64$}
		\label{T=64_4}
	\end{subfigure}
	\begin{subfigure}[b]{0.3\linewidth}
		\includegraphics[width=\linewidth]{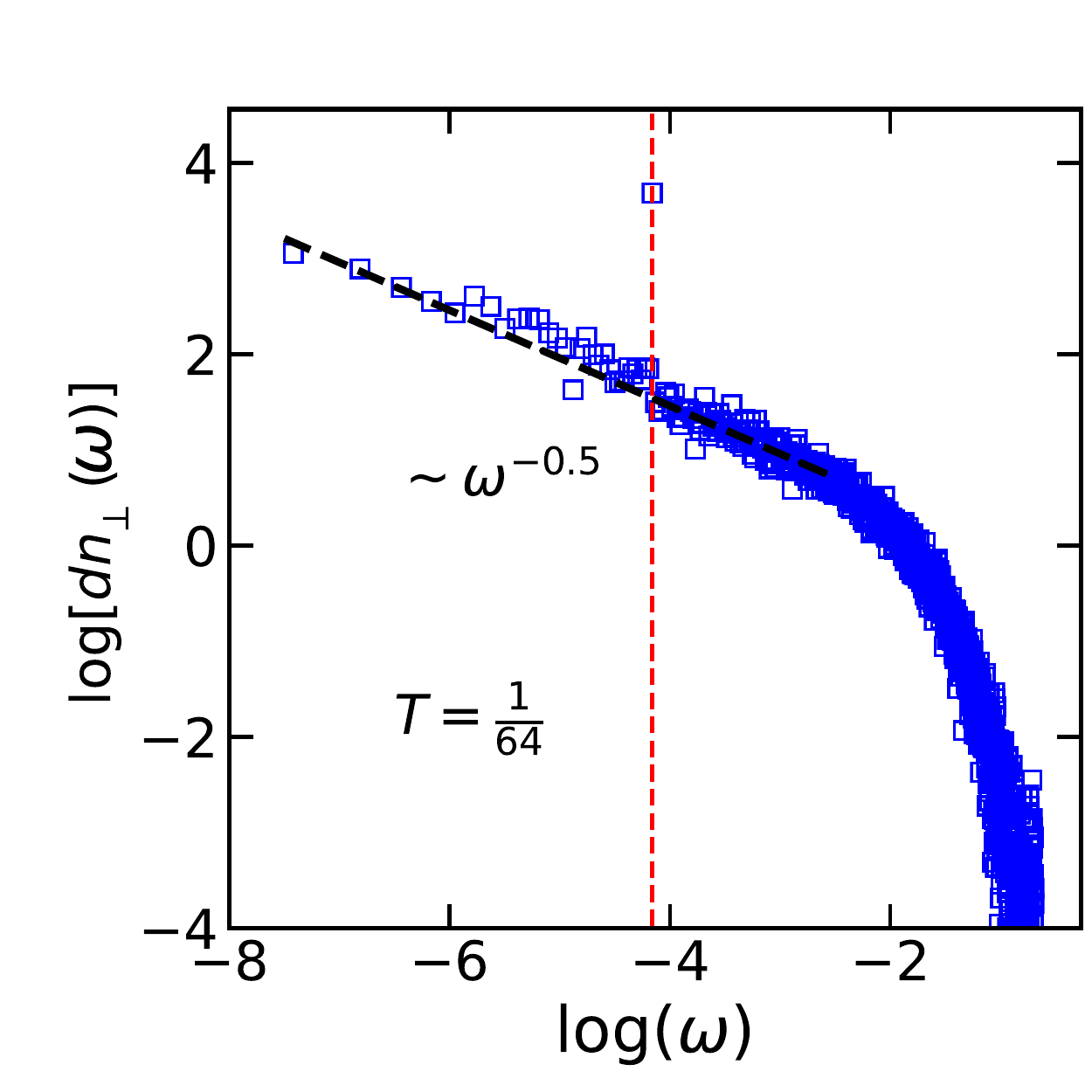}
		\caption{$T=128$}
		\label{T=128_4}
	\end{subfigure}
	\centering
	\caption{PSD for sand dissipation in the perpendicular direction for various $T$ at $L = 256$ lattice size.}
	\label{Fig:PSD_perpendicular}	
\end{figure*}

The other independent quantities of interest are the grain dissipation in the parallel ($dn_{\parallel}$) and perpendicular ($dn_{\perp}$) directions. The ACF of $dn_{\parallel}$ and $dn_{\perp}$ have been shown in Figs.~\ref{Fig:Auto_parallel} and~\ref{Fig:Auto_perpendicular} in SEC.~\ref{SEC:AppendixA}. The decay of ACF for these two quantities is faster than the one for the activity time series. The corresponding PSD's are shown in Figs.~\ref{Fig:PSD_parallel} and~\ref{Fig:PSD_perpendicular} for $dn_{\parallel}$ and $dn_{\perp}$ respectively. It is seen that PSD for $dn_{\parallel}$ reveals the same spectral structure just like the activity field, while for the PSD of $dn_{\perp}$ the position of the peaks are not distinguishable. Actually, by looking at the ACF, we observe that the strength of the oscillations is very weaker for $dn_{\perp}$ than for $dn_{\parallel}$, which is the reason the peaks in PSD of $dn_{\perp}$ is not pronounced and distinguishable. PSD functions show power-law decay for both cases for more than two decades (note that the PSD for the non-oscillating SM $T=2$ can hardly be fitted by a power-law function). The corresponding exponent is fixed ($0.37\pm 0.03$) for $dn_{\parallel}$, while a small change is observed with changing $T$ for $dn_{\perp}$ $\text{exponent} \in [0.37\pm 0.03-0.5\pm 0.03]$. 

\section{conclusion}
This paper was devoted to the sandpile model subjected to a sinusoidal external drive with the period $T$. This problem was analyzed analytically and numerically. Our analytical approach found the Green function in a large $T$ limit, predicting that the avalanches are anisotropic and elongated in the oscillation direction. In particular, we found that the Green function is a multiplication of an exponential term and a modified Bessel function. We considered the problem numerically and showed that in the intermediate $T$ values, the system is found to be in a regime in which the avalanches are elongated in the perpendicular direction with respect to the oscillations. The transition region between these two regimes is identified by measuring the scaling relation between parallel and perpendicular spatial scales. The power spectrum of avalanche size and the grains wasted from the parallel and perpendicular directions are studied, which show power-law behaviour in terms of the frequency with exponents which run with $T$. 

\section{acknowledgment}
The authors would like to thank Deepak Dhar for his constructive help and arguments on the work.

\newpage

\appendix

\section{activity field}\label{SEC:AppendixA}
\begin{figure*}
	\centering
	\includegraphics[width=0.7\linewidth]{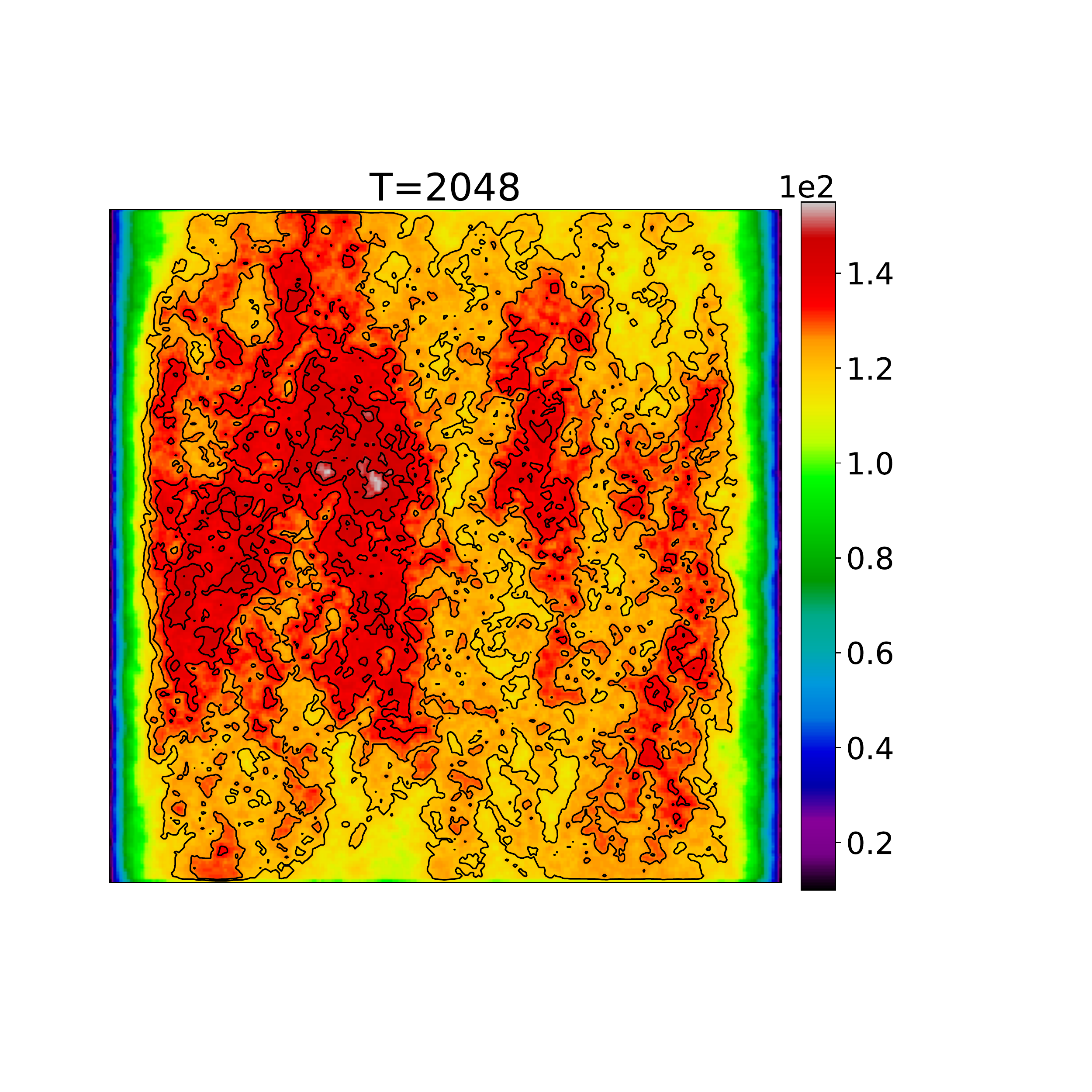}
	\caption{ Total system activity (sum of activities from $t_{\text{reached recurrent}}$ till $t_{\text{finished}}$)  at $T=2048$ on $L=256$ lattice size}%
	\label{total_act_contour_one}	
\end{figure*}

Fig.~\ref{Fig:total_act} shows the activity as a function of for various $T$. It shows stretching in the opposite direction of oscillation for $2<T<256$, and for $T>256$,  stretching in the direction of oscillation. Fig.~\ref{total_act_contour_one} shows $T=2048$ in large frame with contour lines. \\

\begin{figure*}
	\centering
	\includegraphics[width=1.0\linewidth]{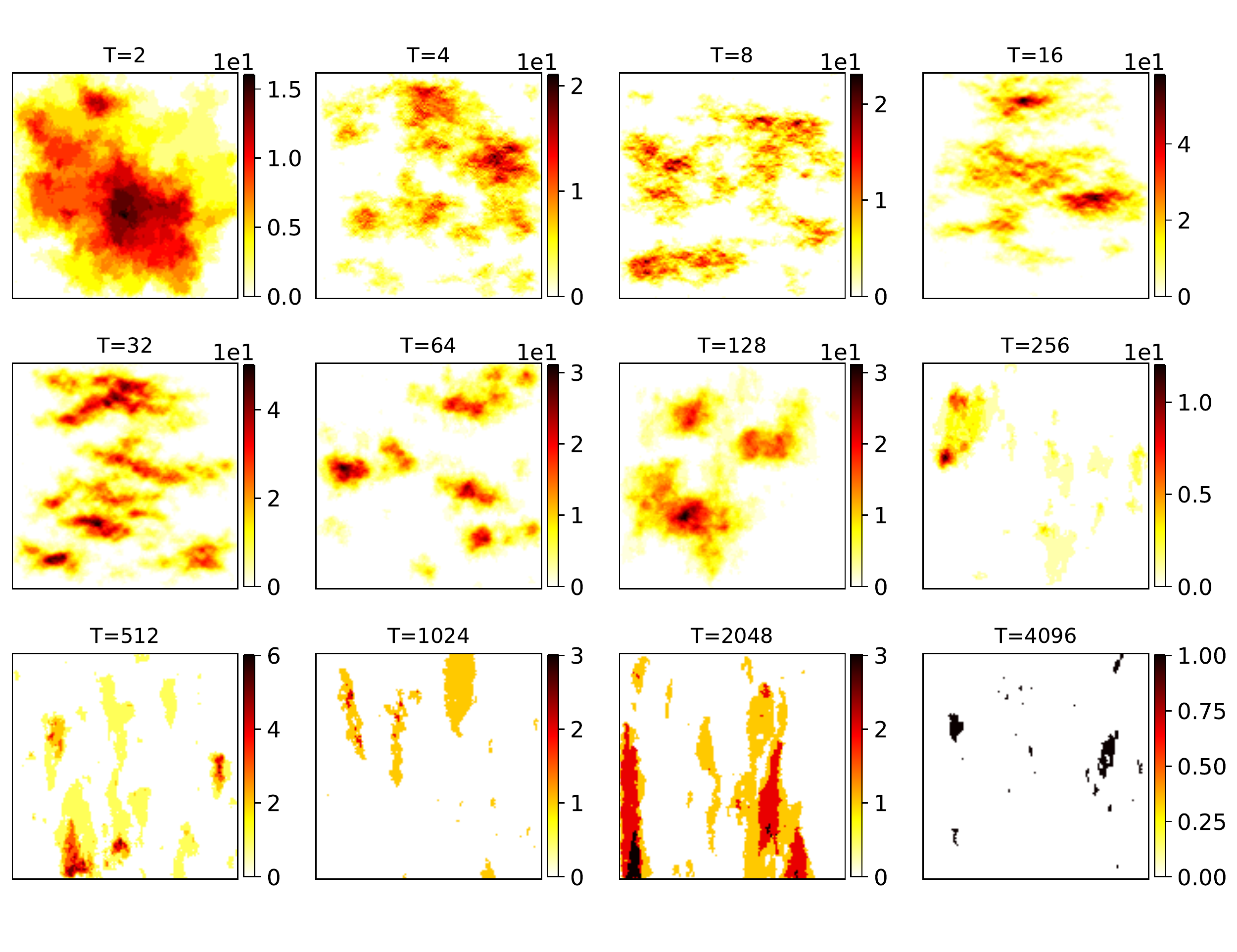}
	\caption{ Finite activity ($t\in[t_{\text{recurrent}}, t_{\text{recurrent}}+1000]$  time step) in recurrent regime for various $T$ on $L=256$ lattice size}	     
	\label{Fig:total_act_contour_1000}	
\end{figure*}

\begin{figure*}
	\begin{subfigure}[b]{0.3\linewidth}
		\includegraphics[width=\linewidth]{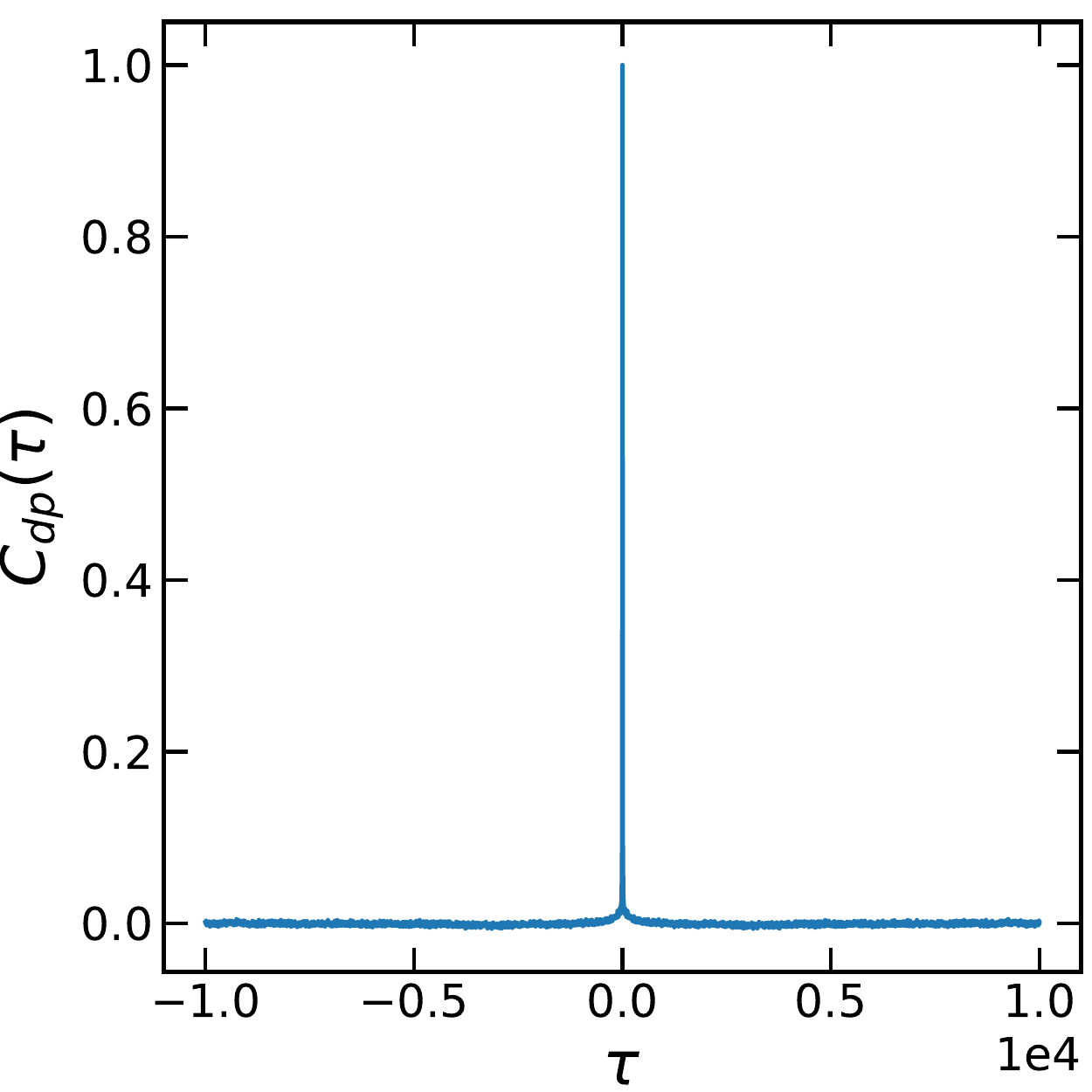}
		\caption{$T=2$}
		\label{T=2_5}
	\end{subfigure}
	\begin{subfigure}[b]{0.3\linewidth}
		\includegraphics[width=\linewidth]{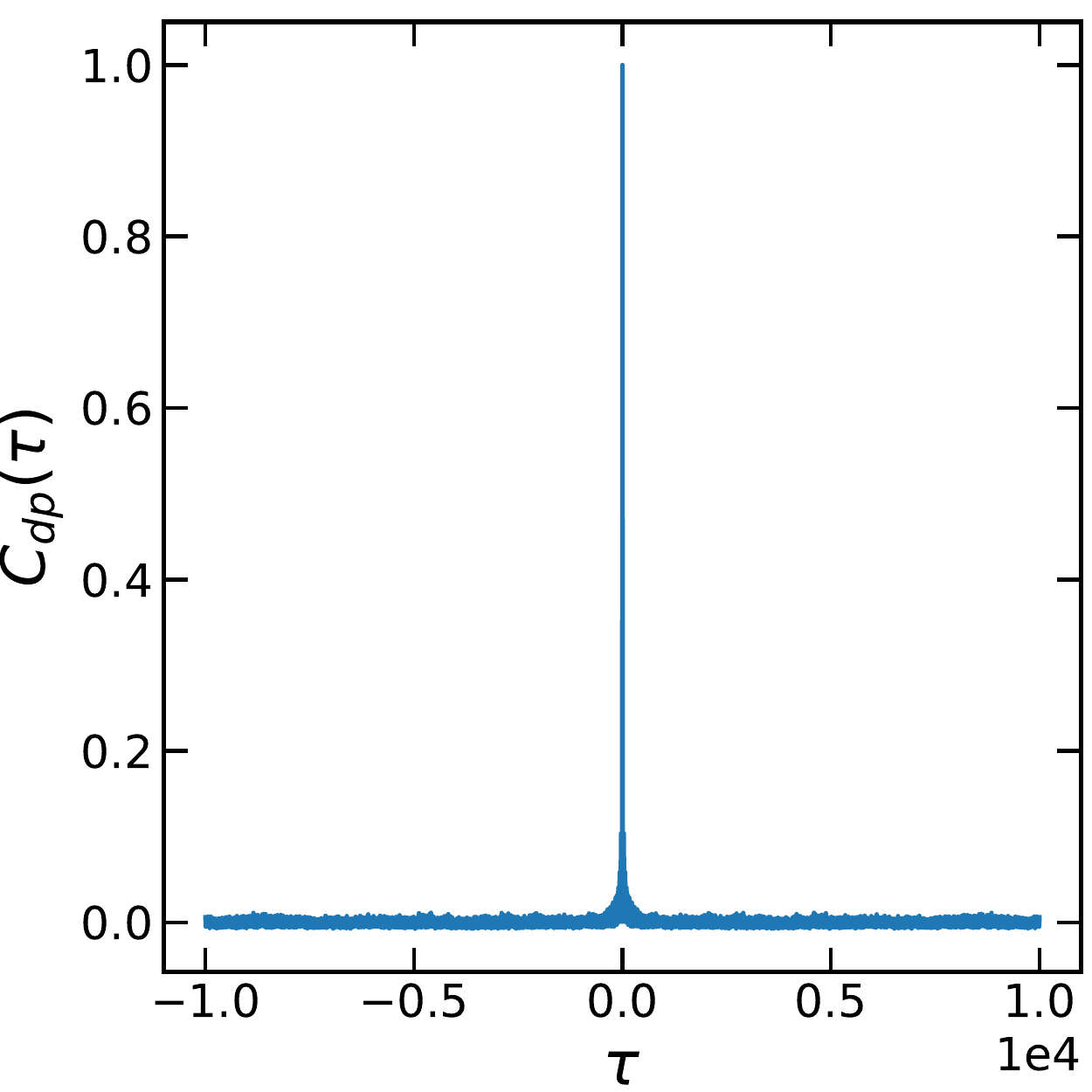}
		\caption{$T=32$}
		\label{T=32_5}
	\end{subfigure}
	\begin{subfigure}[b]{0.3\linewidth}
		\includegraphics[width=\linewidth]{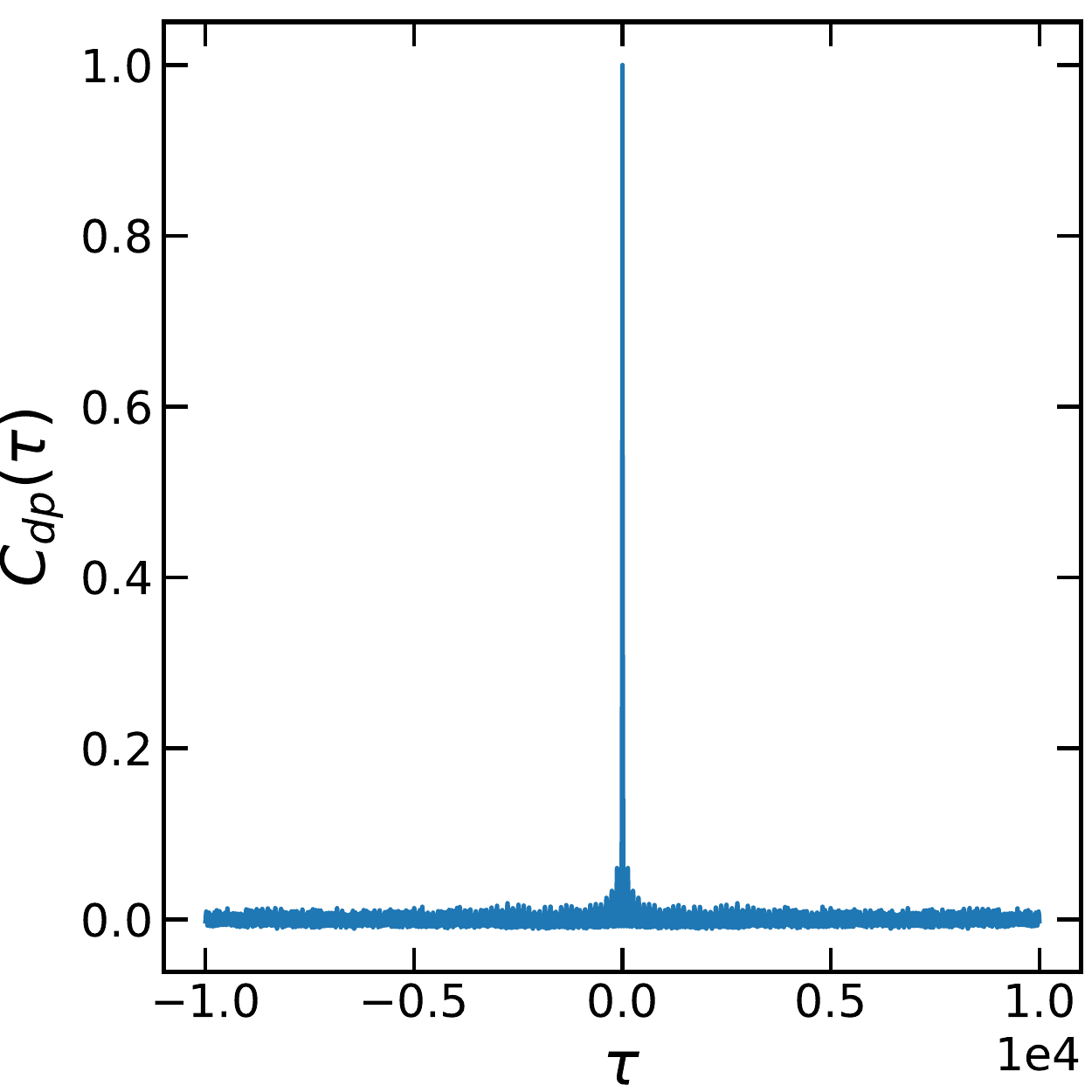}
		\caption{$T=128$}
		\label{T=128_5}
	\end{subfigure}
	\begin{subfigure}[b]{0.3\linewidth}
		\includegraphics[width=\linewidth]{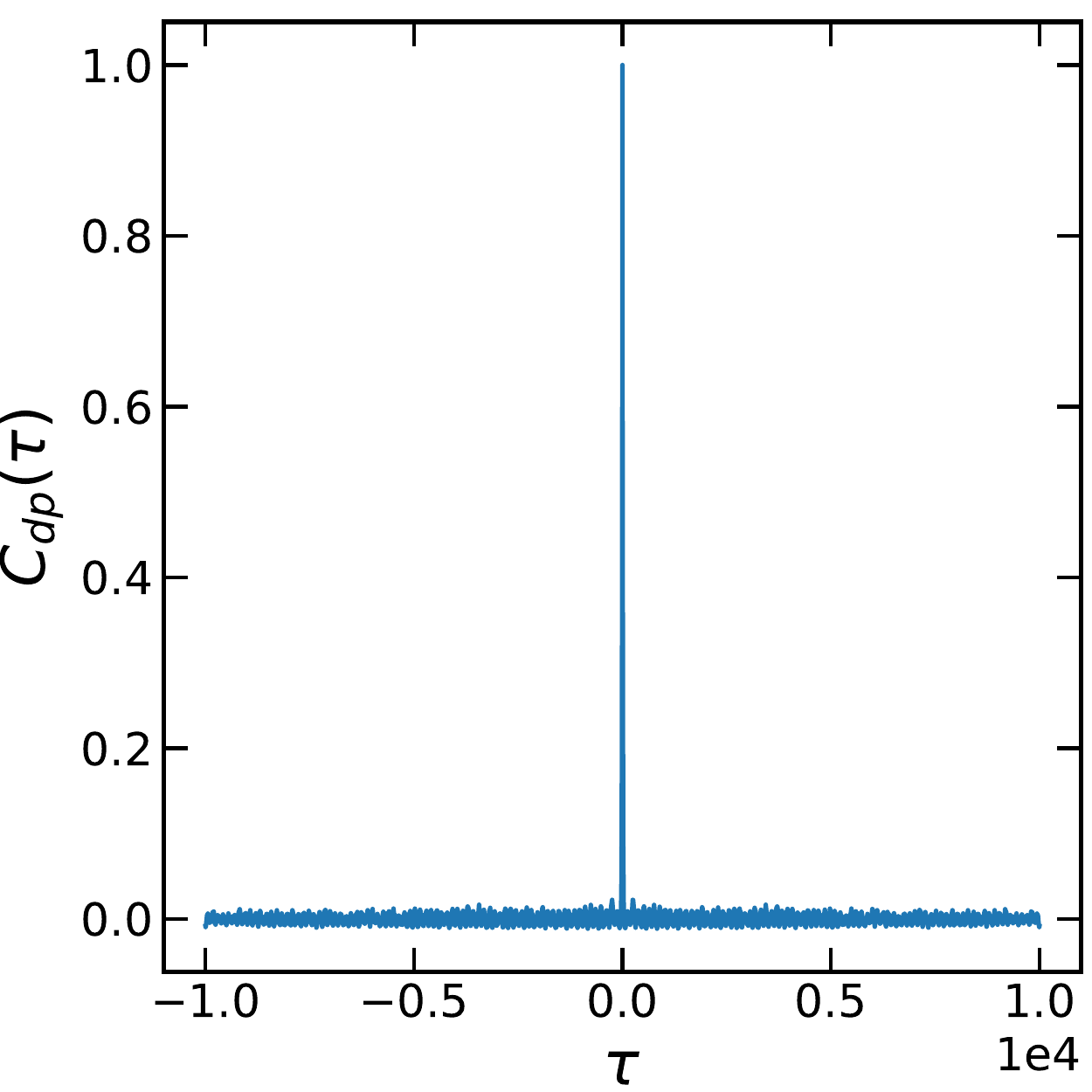}
		\caption{$T=256$}
		\label{T=256_5}
	\end{subfigure}
	\begin{subfigure}[b]{0.3\linewidth}
		\includegraphics[width=\linewidth]{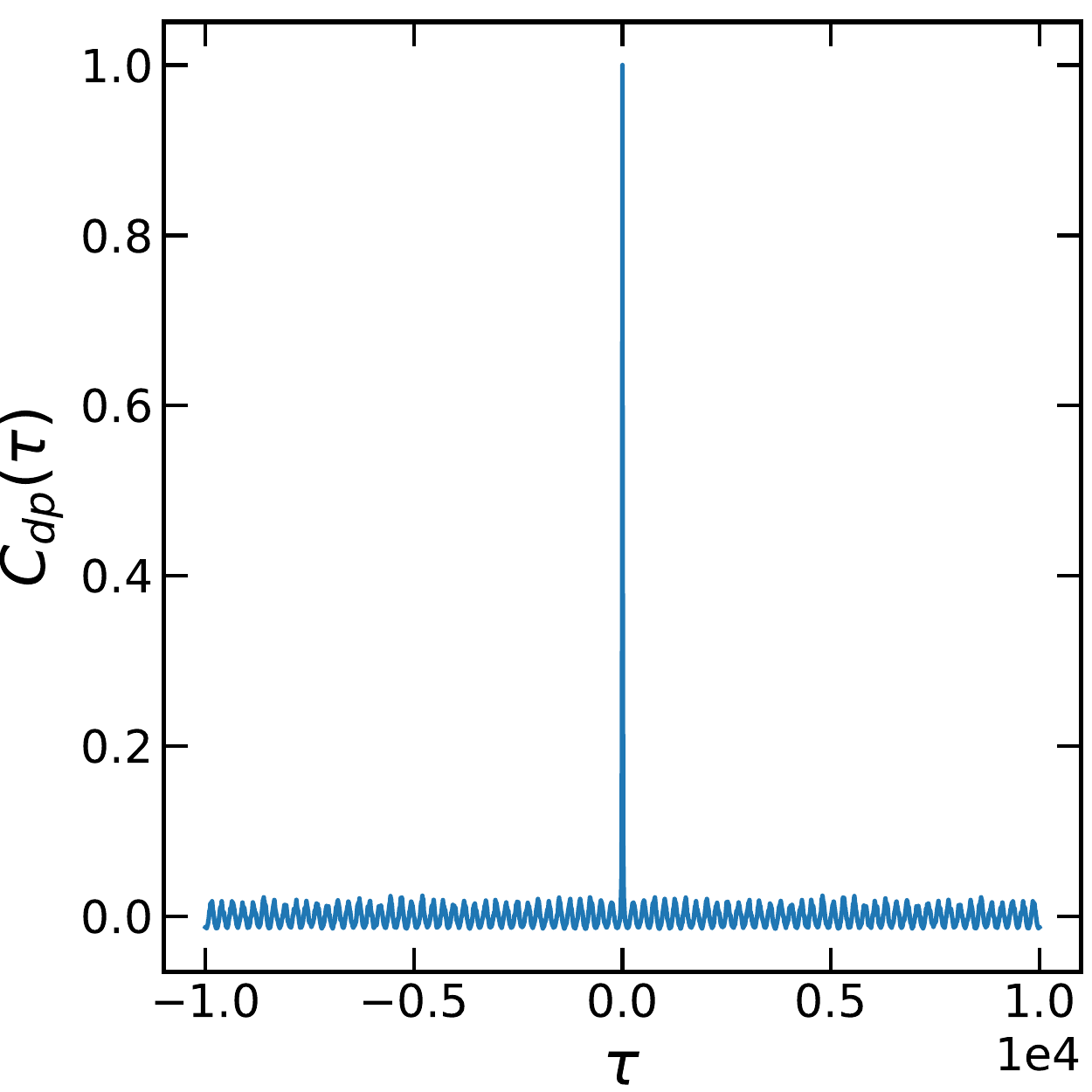}
		\caption{$T=512$}
		\label{T=512_5}
	\end{subfigure}
	\begin{subfigure}[b]{0.3\linewidth}
		\includegraphics[width=\linewidth]{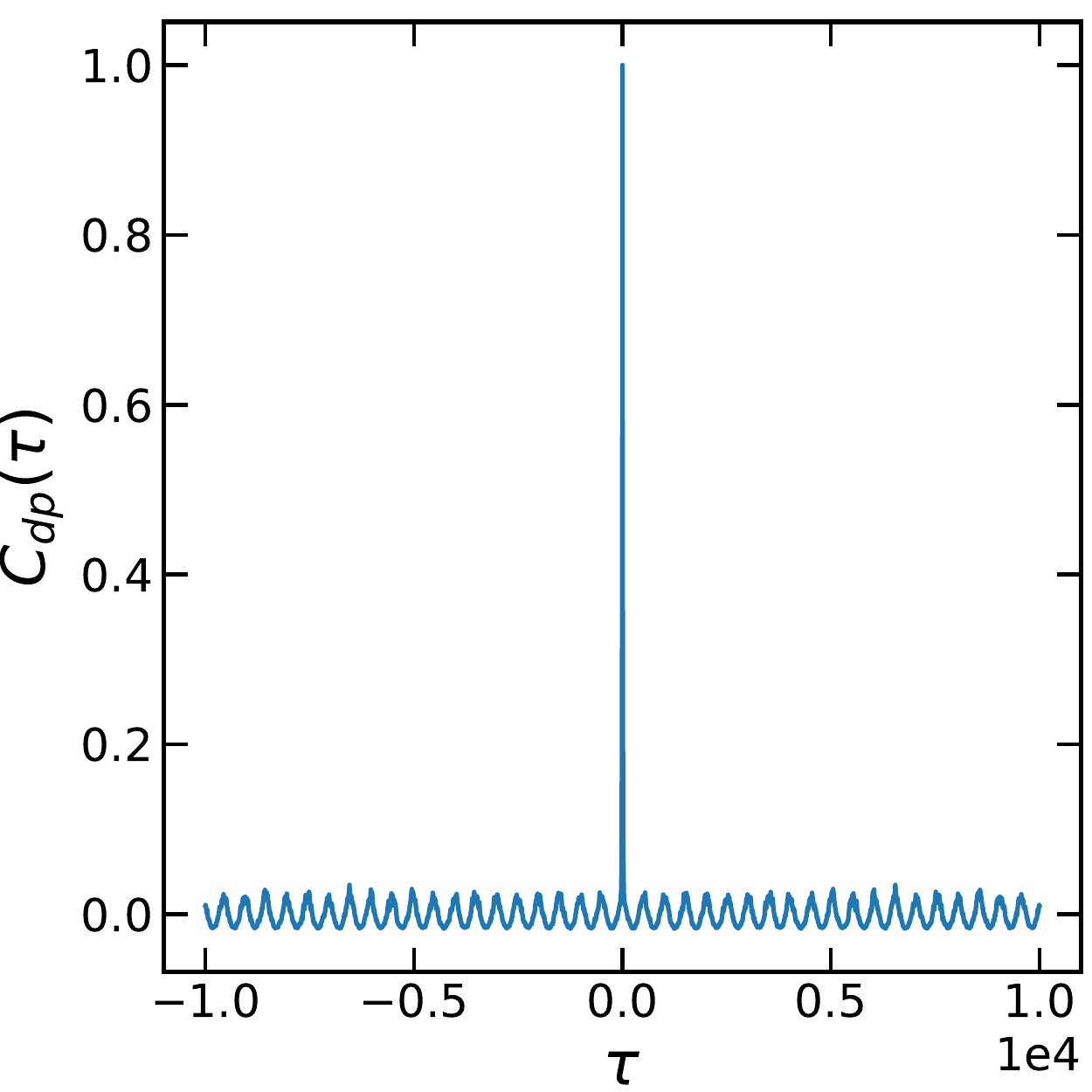}
		\caption{$T=1024$}
		\label{T=1024_5}
	\end{subfigure}
	\centering
	\caption{ACF of sand dissipation in the parallel direction for various $T$ at $L = 256$ lattice size.}
	\label{Fig:Auto_parallel}	
\end{figure*}

\begin{figure*}
	\begin{subfigure}[b]{0.3\linewidth}
		\includegraphics[width=\linewidth]{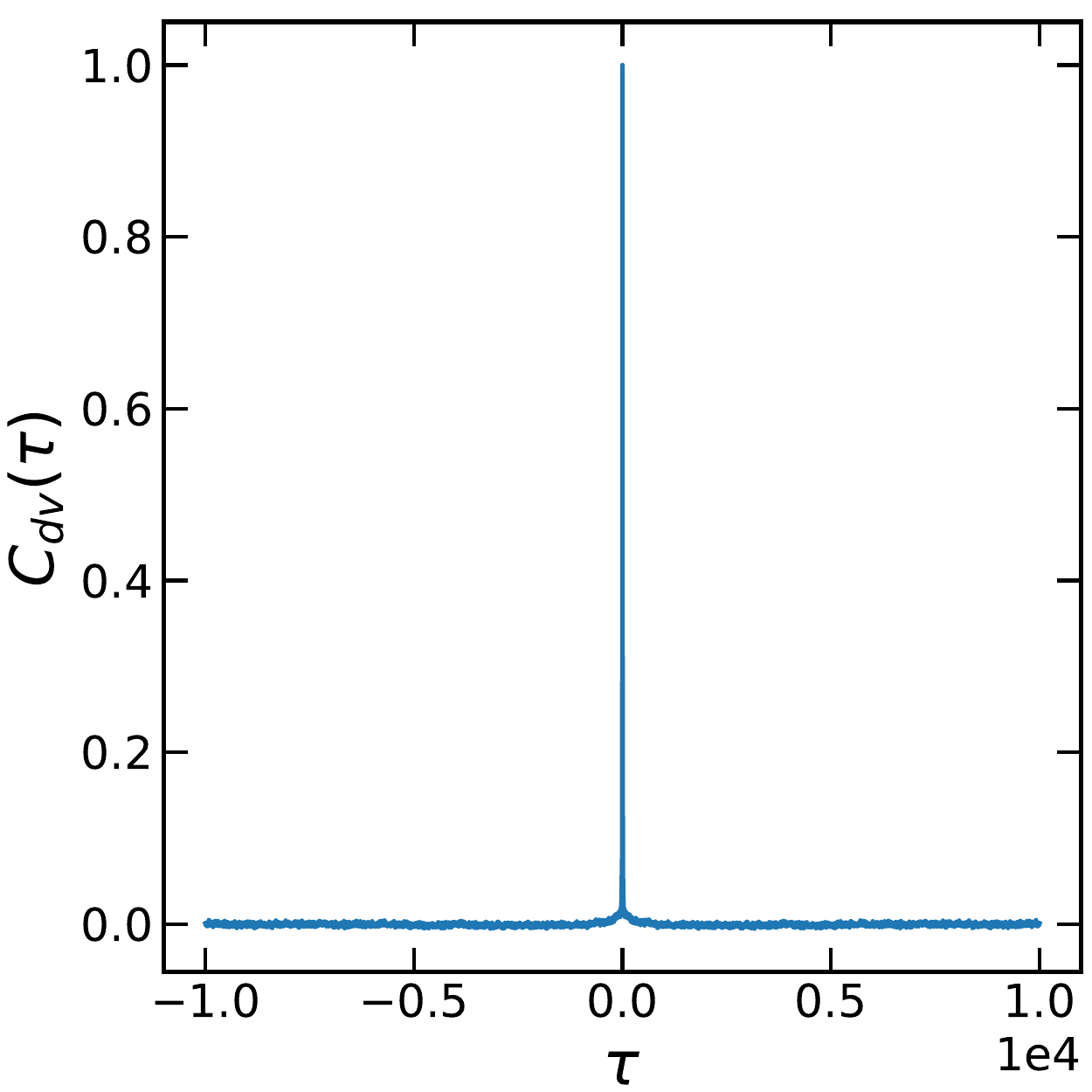}
		\caption{$T=2$}
		\label{T=2_3}
	\end{subfigure}
	\begin{subfigure}[b]{0.3\linewidth}
		\includegraphics[width=\linewidth]{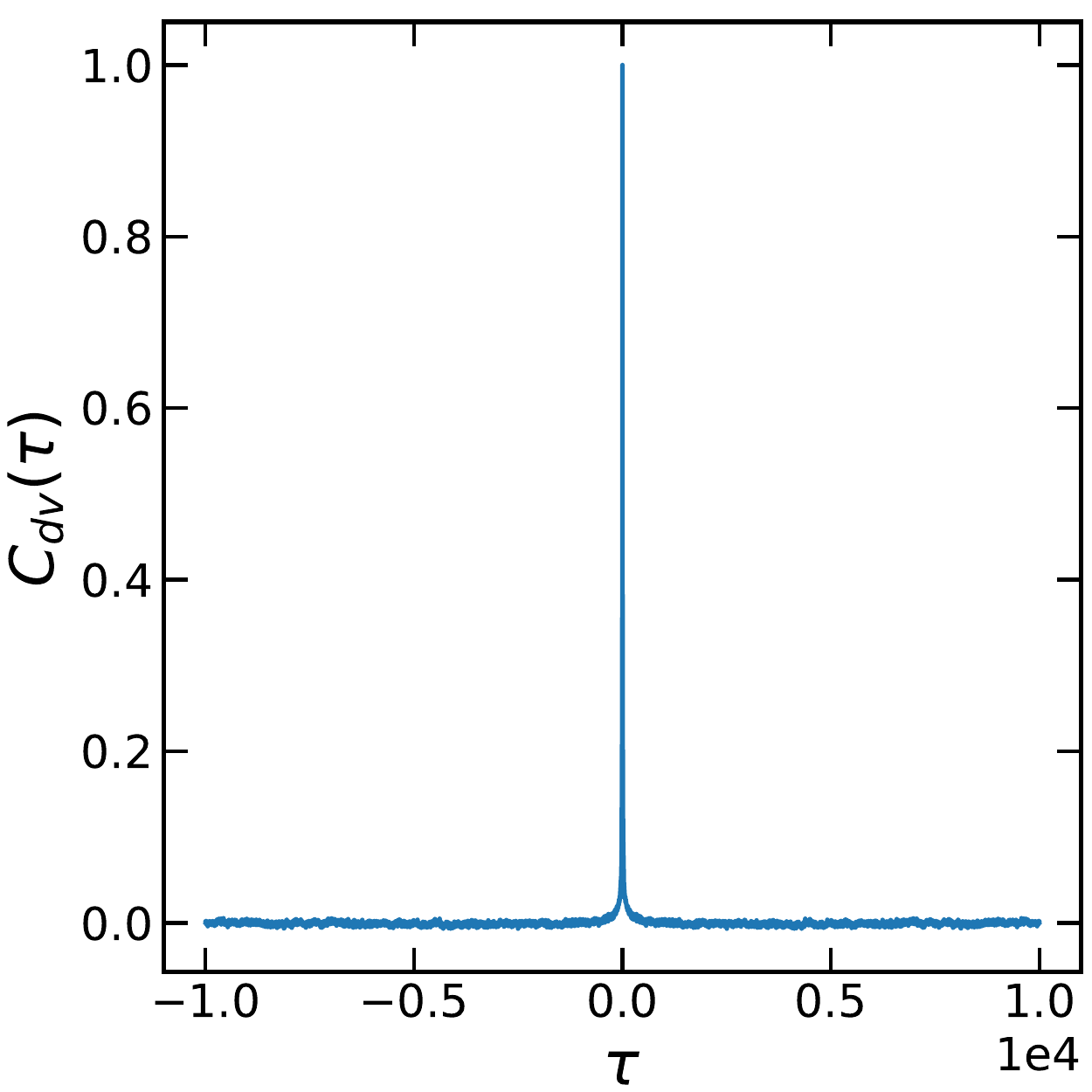}
		\caption{$T=32$}
		\label{T=32_3}
	\end{subfigure}
	\begin{subfigure}[b]{0.3\linewidth}
		\includegraphics[width=\linewidth]{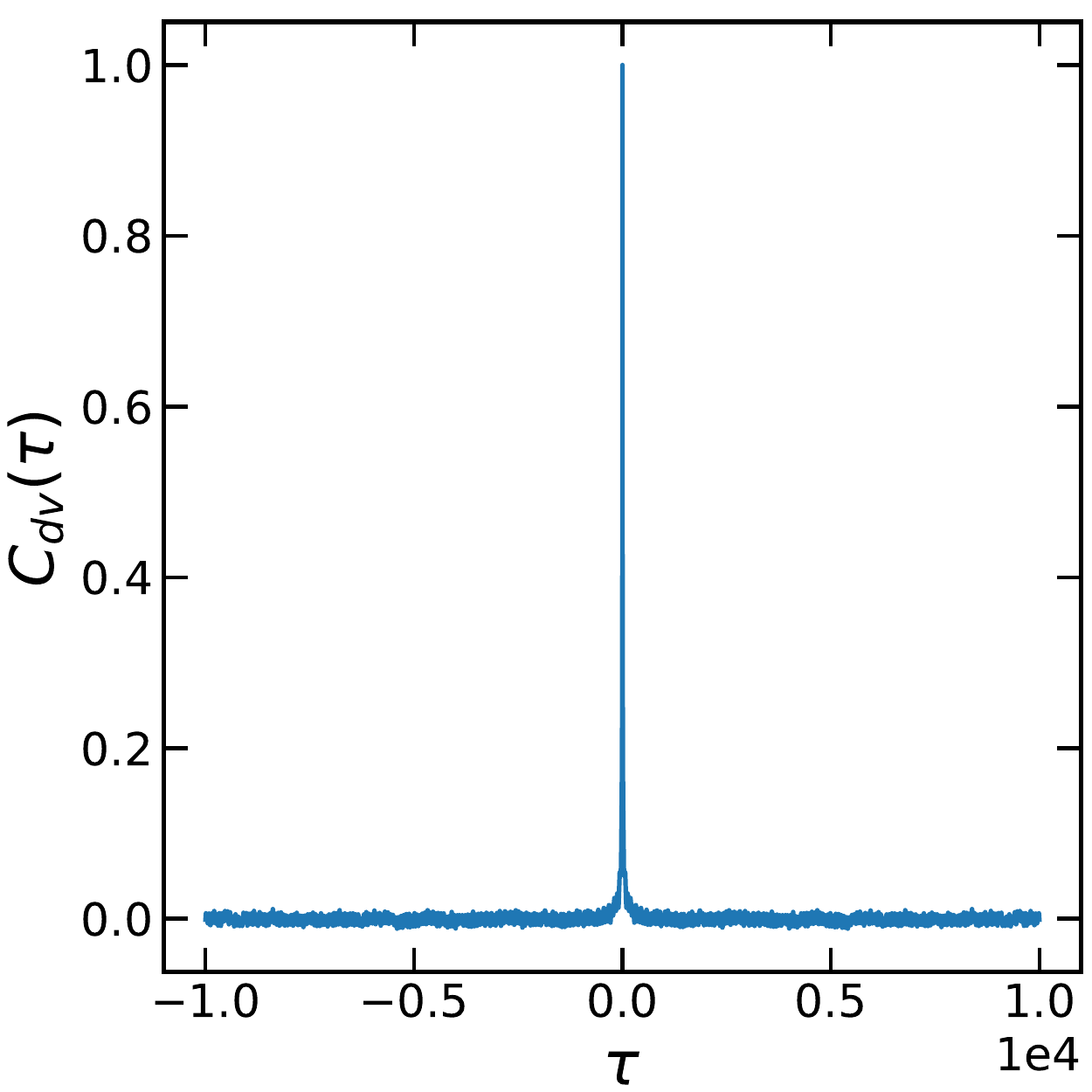}
		\caption{$T=128$}
		\label{T=128_3}
	\end{subfigure}
	\begin{subfigure}[b]{0.3\linewidth}
		\includegraphics[width=\linewidth]{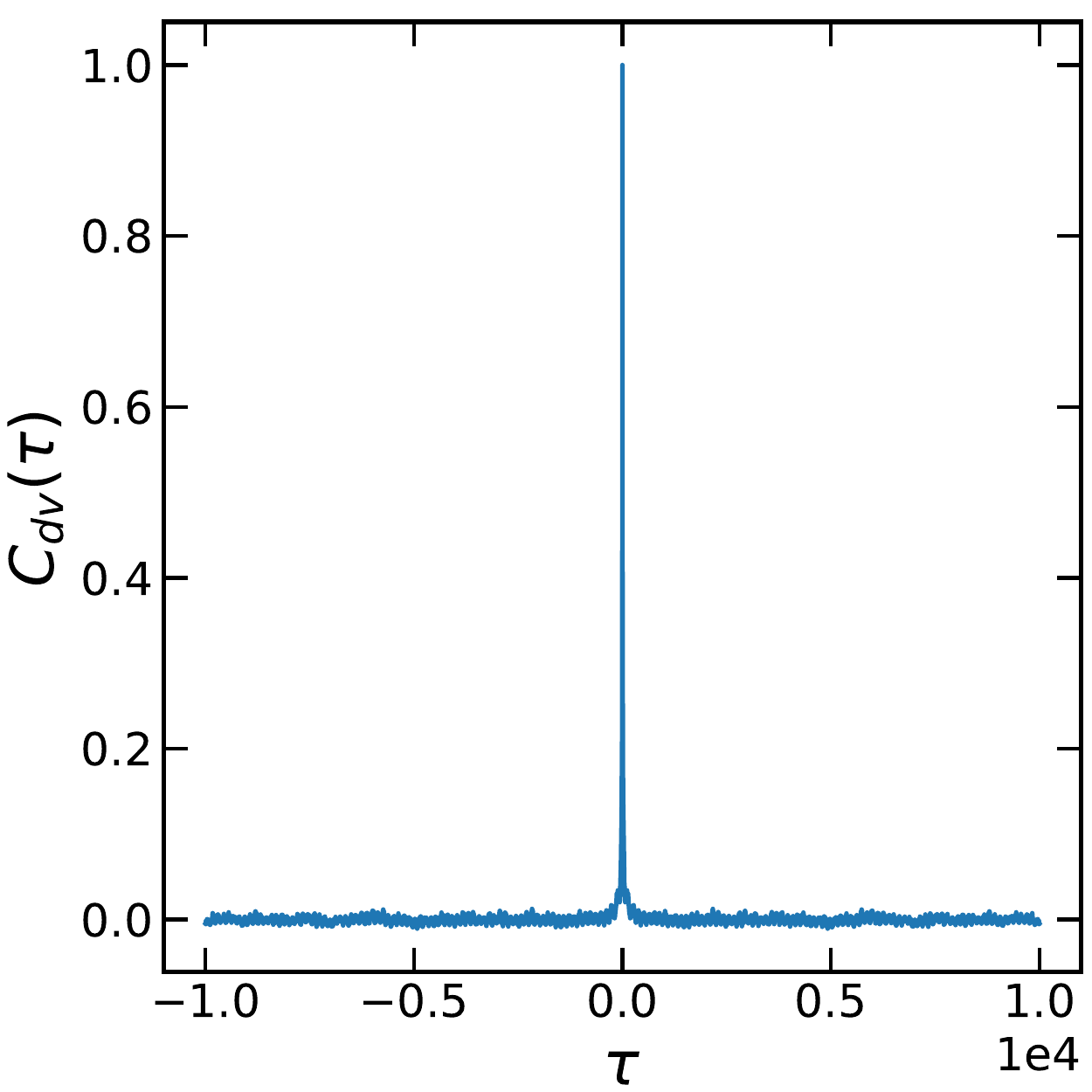}
		\caption{$T=256$}
		\label{T=256_3}
	\end{subfigure}
	\begin{subfigure}[b]{0.3\linewidth}
		\includegraphics[width=\linewidth]{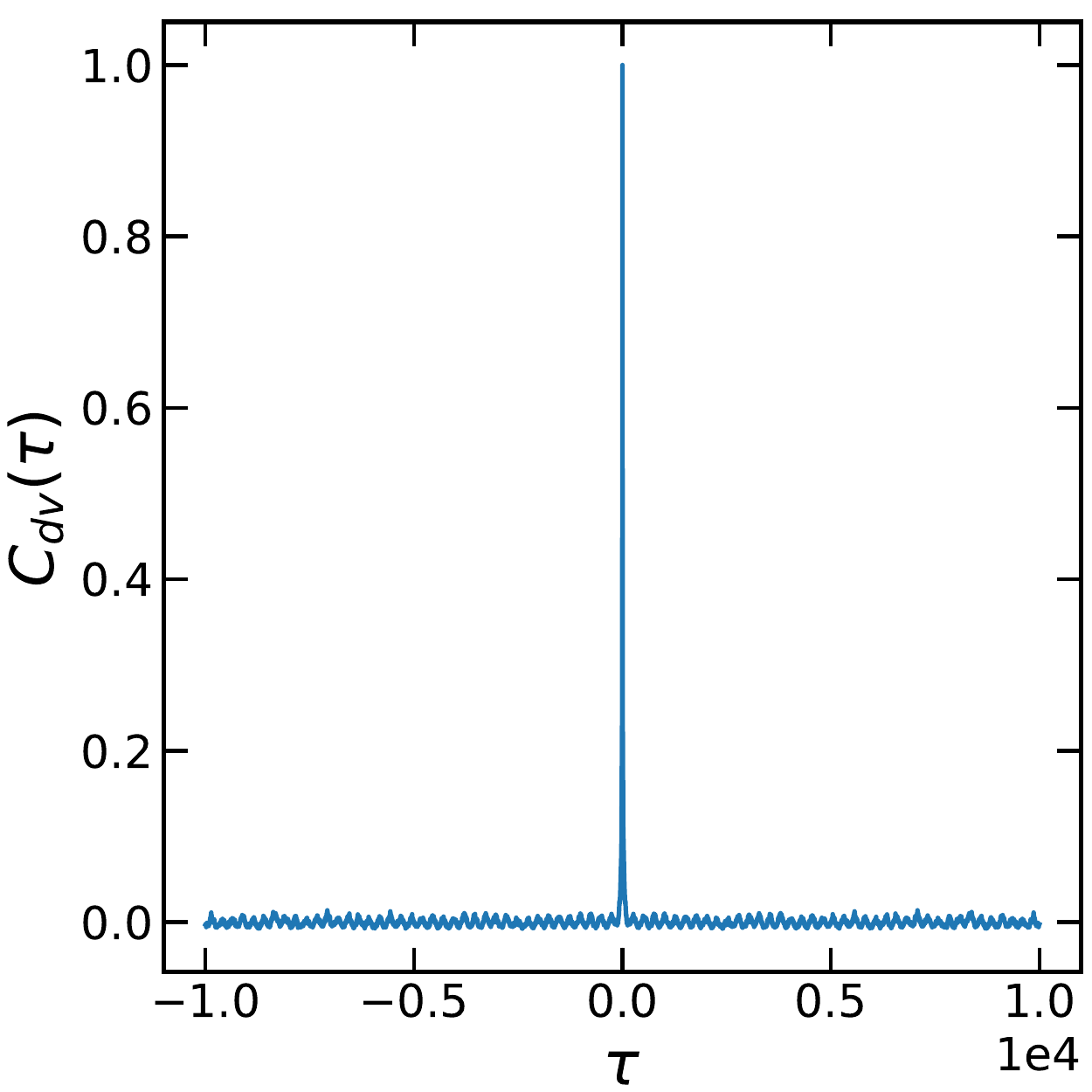}
		\caption{$T=512$}
		\label{T=512_3}
	\end{subfigure}
	\begin{subfigure}[b]{0.3\linewidth}
		\includegraphics[width=\linewidth]{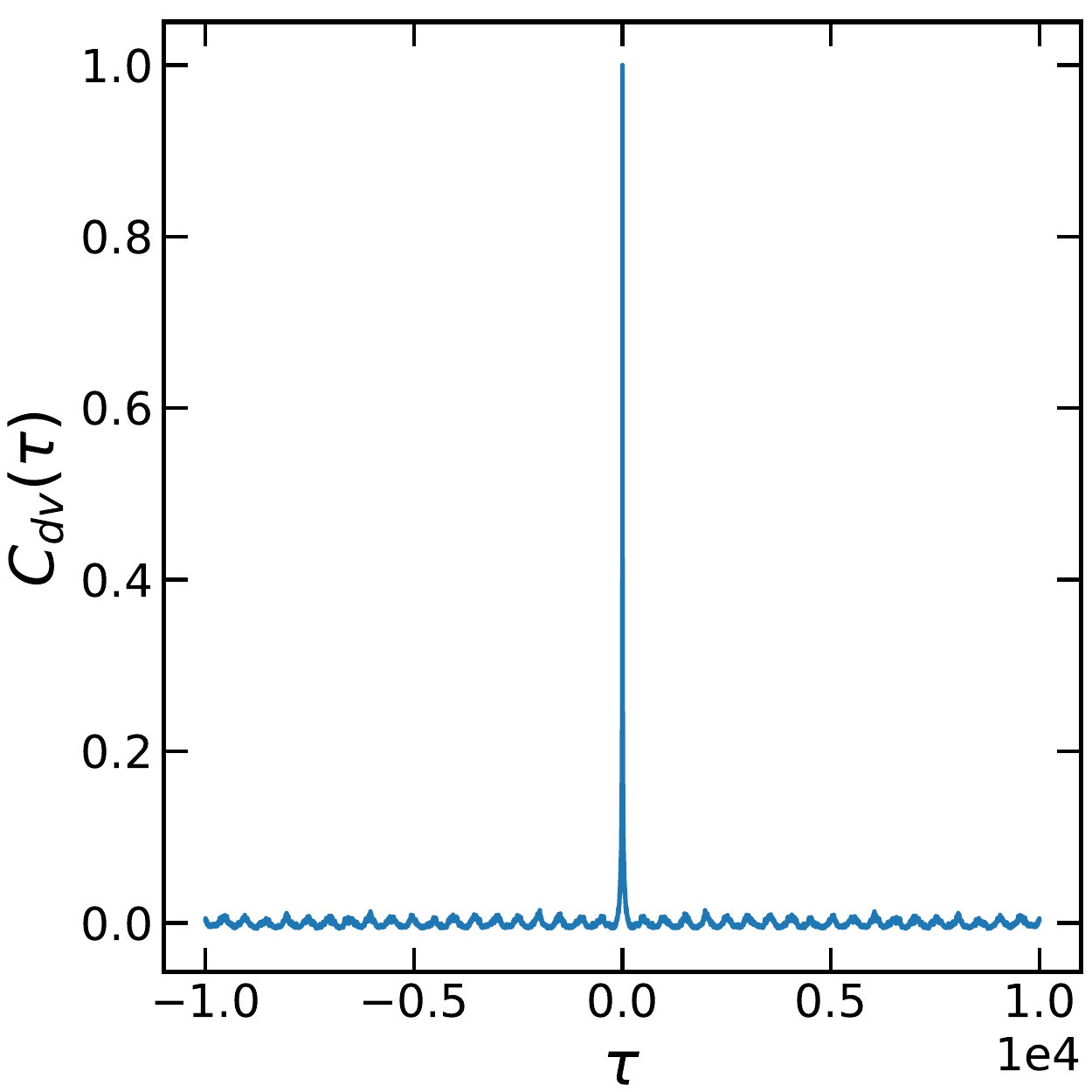}
		\caption{$T=1024$}
		\label{T=1024_3}
	\end{subfigure}
	\centering
	\caption{ACF of sand dissipation in the perpendicular direction of oscillation for various $T$ at $L = 256$ lattice size.}
	\label{Fig:Auto_perpendicular}	
\end{figure*}

Fig.~\ref{Fig:total_act_contour_1000} shows the system activity in a finite time step, i.e., $t\in[t_{\text{recurrent}}, t_{\text{recurrent}}+t_{\text{finit}}]$. This figure shows the toppling number in recurrent regime for $t=1000$ time steps. This figure shows system activity decreases for $T\geq256$.
Figs.~\ref{Fig:total_act_contour1} and \ref{Fig:total_act_contour2} show finite system activity for $T=2$ and $T=64$ with contour line in large scale., and also Fig.~\ref{Fig:total_act_contour3} show $T=2048$  for $t= 10000$ time steps in recurrent regime. It shows this fact that the activity for large $T$ is significantly low.

\bibliography{refs}

\end{document}